\documentclass[aps,pof,reprint,groupedaddress]{revtex4-1}
\pdfoutput=1
\usepackage{graphicx}
\usepackage{epstopdf, epsfig,setspace}
\usepackage{longtable}
\usepackage{import}
\usepackage{color}
\usepackage{todonotes}
\usepackage{amsmath}
\usepackage{amssymb}
\usepackage{natbib}
\usepackage[noabbrev]{cleveref}
\newcommand{\scalefig}{1.05}

\newcommand{\dd}[1]{\mathrm{d}#1}

\newcommand{\Rey}{Re}

\begin{document}
\title{Turbulent skin-friction reduction by wavy surfaces}
\author{Sacha Ghebali}
\email{s.ghebali14@imperial.ac.uk}
\author{Sergei I. Chernyshenko}
\email{s.chernyshenko@imperial.ac.uk}
\author{Michael A. Leschziner}
\email{mike.leschziner@imperial.ac.uk}
\affiliation{Department of Aeronautics, Imperial College London,
South Kensington Campus, London SW7 2AZ, United Kingdom}

\date{\today}

\begin{abstract}
Direct numerical simulations of fully-developed turbulent channel flows with wavy walls are undertaken. The wavy walls, skewed with respect to the mean flow direction, are introduced as a means of emulating a Spatial Stokes Layer (SSL) induced by in-plane wall motion. The transverse shear strain above the wavy wall is shown to be similar to that of a SSL, thereby affecting the turbulent flow, and leading to a reduction in the turbulent skin-friction drag. The pressure- and friction-drag levels are carefully quantified for various flow configurations, exhibiting a combined maximum overall-drag reduction of about 0.5\%. The friction-drag reduction is shown to behave approximately quadratically for small wave slopes and then linearly for higher slopes, whilst the pressure-drag penalty increases quadratically.
Unlike in the SSL case, there is a region of increased turbulence production over a portion of the wall, above the leeward side of the wave, thus giving rise to a local increase in dissipation.
The transverse shear-strain layer is shown to be approximately Reynolds-number independent when the wave geometry is scaled in wall units.
\end{abstract}

\maketitle

\section{Introduction}

In the context of greener and more cost-effective aviation, industrial and academic researchers have proposed and studied a wide range of control methods mainly over the past three decades.
Unfortunately, hardly any offers realistic prospects of being implemented in practice.
This applies, in particular, to active control schemes, despite some successful implementations of active devices in laboratory tests.

Among passive concepts, the use of riblets was originally inspired by the narrow grooves observed on sharks' placoid scales.
Although the effectiveness of the dermal denticles of the shark has been questioned by \citet{Boomsma2016}, the use of optimally-chosen longitudinal grooves, aligned with the main flow direction, has been shown to be capable of reducing the turbulent skin-friction drag by levels of order 5--10\% \citep{Choi1993,Bechert1989,Bechert1997,Garcia-Mayoral2011}.
However, a practical, cost-effective implementation has yet to be achieved, mostly hindered by the small optimal spacing required (about 15$\mu$m in cruise-speed conditions) and stringent tolerances on the sharpness of the crests.
More complex variants, such as sinusoidal riblets, were studied in \citep{Peet2009b,Kramer2010,BannierPhD}, but despite attempts to optimise the geometry, \citet{BannierPhD} showed that conventional (straight) riblets appear to be as effective.

On the active-control front, based upon the work of \citet{Jung1992} on the drag-reducing properties of transverse wall oscillations, it has been established, computationally, that imparting streamwise-modulated, spanwise in-plane wall motions of the form of a travelling wave ${w_w(x,t) = A \sin(2\pi\, x/\lambda_x - \omega t)}$, giving rise to a `Generalised Stokes Layer' (GSL), results in gross friction-drag-reduction levels of up to about 45\% \citep{Quadrio2009,Quadrio2010} at low Reynolds numbers, the effectiveness being observed to reduce at higher Reynolds numbers \citep{Touber2012,Hurst2014,Gatti2016}.
An experimental confirmation of the numerical results  for the travelling wave was undertaken by \citet{Auteri2010} in pipe flows, for which drag-reduction levels of up to 33\% were achieved, while \citet{Bird2015} designed a Kagome-lattice-based actuator for boundary layers, achieving a drag-reduction level of around 10\%.
However, it is very challenging to implement the latter method in a physical laboratory, let alone in practice.

\begin{figure}\centering
\vspace{10pt}
\includegraphics[width=\linewidth]{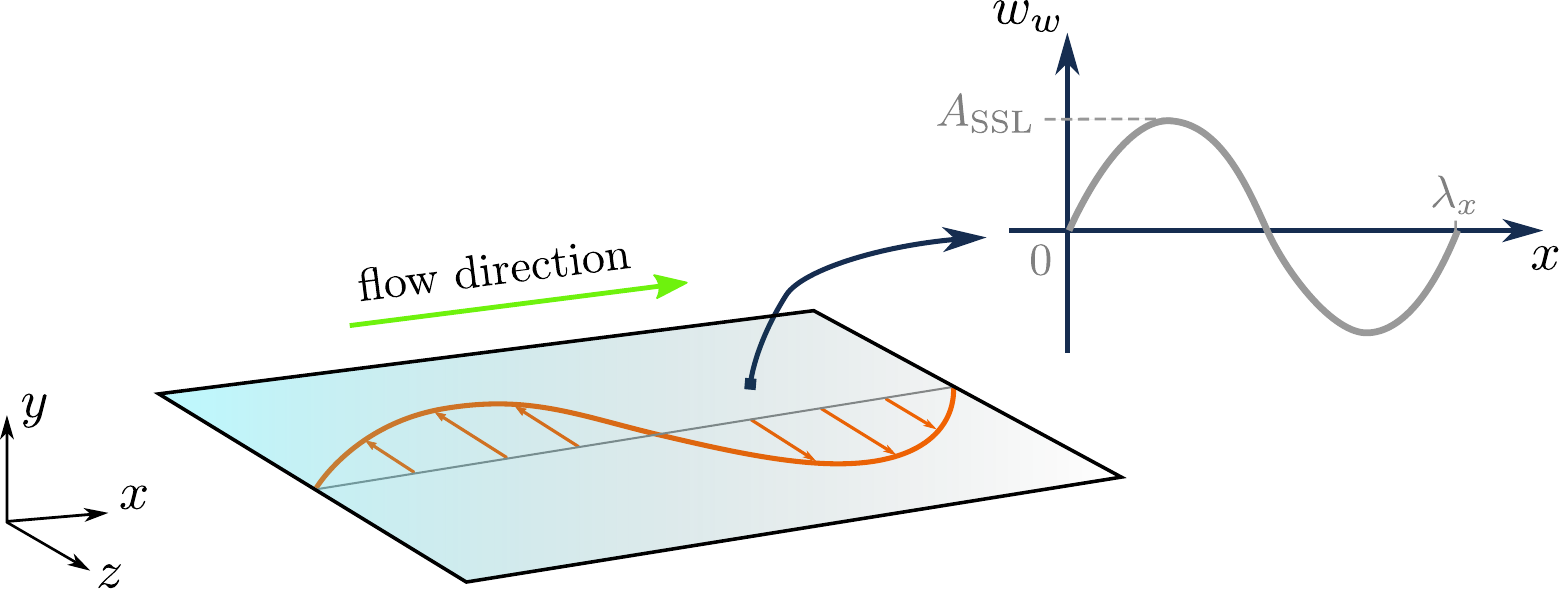}
\caption{Schematic of the in-plane wall motion imparted in the case of a Spatial Stokes Layer (SSL).}
\label{fig:SSL}
\end{figure}

A particular case of the GSL is the Spatial Stokes Layer (SSL), consisting of a standing wave ${w_w=A_\text{SSL}\sin(2\pi\,x/\lambda_x)}$, as shown in \cref{fig:SSL}.
This method was studied by \citet{Viotti2009} by means of DNS for various forcing amplitudes $A_\text{SSL}$ and wavelengths $\lambda_x$.
The maximum net-energy savings of 23\%, achieved as a result of \citeauthor{Viotti2009}'s exploration at $\Rey_\tau =O(200)$, was for a forcing wavelength of $\lambda_x^+ = O(1000)$ that is, about two orders of magnitude larger than the optimal dimension of riblets.
Thus, while the resulting control method is still active, it is steady, and based on geometric dimensions that, for an entirely passive device, would be compatible with a practical implementation.

\begin{figure*}\centering
\includegraphics[width=.8\textwidth]{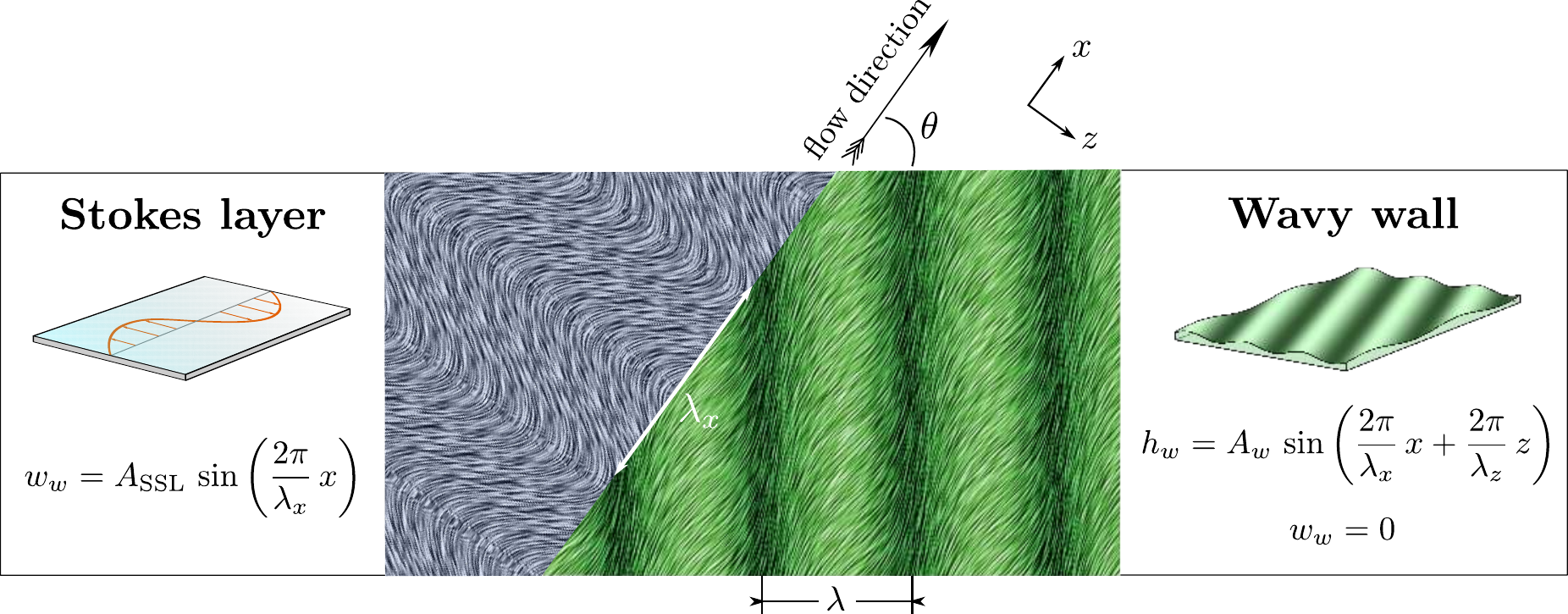}
\caption{Emulation of the forcing. Visualisation of mean streaklines close to the wall for SSL (left) and wavy wall (right).  The background is coloured by the norm of the velocity vector.}
\label{fig:flowvisu}
\end{figure*}

In an effort to address the need for practical control methods, the present research examines a passive means of emulating spanwise in-plane wall motions, as proposed in \cite{Chernyshenko2013}.
The geometry considered is a solid wavy wall, with the troughs and crests skewed with respect to the main flow direction.
The presence of the skewed wavy wall creates a spanwise pressure gradient that forces the flow in the spanwise direction, thus generating an alternating spanwise motion, as shown in \cref{fig:flowvisu}.
In contrast to the SSL, where the wall is actuated, the velocity has to vanish at the solid wall, so that the spanwise forcing can obviously not be faithfully emulated.
However, the premise is that the wavy geometry will generate a spanwise shear strain, somewhat away from the wall, that will weaken turbulence in a similar manner to that effected by the SSL.
Such a passive device would benefit from the favourable actuation characteristics of the SSL (large wavelength), resulting in a practical solution, from a manufacturing and maintenance standpoint.

The present study will focus on selected direct numerical simulations of turbulent wavy-channel flows with the aim of examining the degree of Stokes-layer emulation achieved, and the degree to which the drag is reduced relative the plane channel.
As part of this study, some major similarities and differences between the flow arising from in-plane wall motions and that past a wavy wall, as shown in \cref{fig:flowvisu}, will be brought to light, including the impact on the near-wall turbulence.


\section{Methodology}
\subsection{Overall strategy}
An obvious problem is posed by the absence of any guidance on which combination of geometric parameters offers the promise of maximum drag reduction.
An exploration of the three-dimensional parameter space (wave height, wavelength, and flow angle) by a `carpet-bombing' strategy, or classical formal optimisation strategy, is not tenable on cost grounds, especially because of the tight resolution requirements needed for an accurate prediction of the drag increase/decrease margin.
For this reason, a preliminary low-order study was undertaken by \citet{Chernyshenko2013} to narrow down the exploration range within which the drag reduction might be maximised.
A preliminary study of this type was undertaken by \citeauthor{Chernyshenko2013} who found an estimate for the streamwise-projected wavelength of the wave ${\lambda_x^+ \approx 1500}$, and the flow angle ${\theta\approx 52^\circ}$, but who did not provide an estimate for the height of the wave.
Rather, a condition was given for the wave height, subject to the amplitude of the forcing of the emulated SSL $A_\text{SSL}^+=2$.

The present strategy was initiated with the configuration given in \citet{Chernyshenko2013}. The wave height was chosen in order to approximately satisfy the above-mentioned condition on the emulated SSL.
Other configurations, a selection of which will be presented below, were later simulated, and the exploration was mainly undertaken by trial an error.

\subsection{Computational simulations}
Direct Numerical Simulations are performed using an in-house code, that features collocated-variable storage, second-order spatial approximations implemented within a finite-volume, body-fitted mesh.
The equations are explicitly integrated in time by a third-order gear-like scheme, described in~\cite{Fishpool2009}.
In the fractional-step procedure, the non-solenoidal intermediate velocity field is projected onto the solenoidal space by solving a pressure--Poisson equation.
The latter is solved by Successive Line Over-Relaxation \citep[p.~510]{Hirsch2007}, the convergence of which is accelerated using a multigrid algorithm by \citet{Lien1993}. The multigrid iterations within any time step are terminated when a convergence criterion, based on the RMS of the mass residuals, made non-dimensional using the fluid density, bulk velocity, and channel half-height, is met. A typical value used for this criterion is $10^{-10}$.
Stable velocity--pressure coupling is ensured by use of the Rhie-and-Chow interpolation \citep{Rhie1983}, preventing odd-even oscillations.

The code has been thoroughly verified and validated. 
A verification of the spatial accuracy of the code was undertaken via the Method of Manufactured Solutions \citep{Roache1997,Roache1998,Roache2002,Roy2004,Salari2000}, which indicated a second-order spatial accuracy for the velocity and pressure fields.
The manufactured solution was implemented in a channel with lower and upper walls being wavy and flat, respectively.
Validation was performed by independently reproducing the results of flow solutions documented in existing databases. Thus, results were obtained for a turbulent flow past a wavy wall and compared to the experimental and DNS data by \citet{Maass1996}, provided in the ERCOFTAC Classic Database (case~77). Very good agreement was found for all the statistical quantities available, including the velocity, Reynolds stresses, and pressure field.

\subsection{Spatial discretisation of the problem}
\subsubsection{Simulation of a wavy channel}
The simulations were performed using surface-conformal meshing.
The wavy geometry was created by adding an increment $h_w(x,y,z)$ to the wall-normal cell coordinates of a plane channel of half-height $h$, with the walls located at $y=\pm h$.

A number of grid configurations have been simulated, with the characteristics of the plane-channel mesh ($h_w = 0$) listed in \cref{tab:grids}, where all quantities are scaled by reference to the target friction Reynolds number.
The labels G1 to G6 will be used later to identify these cases.

\begin{figure*}
\centering
\includegraphics[scale=\scalefig]{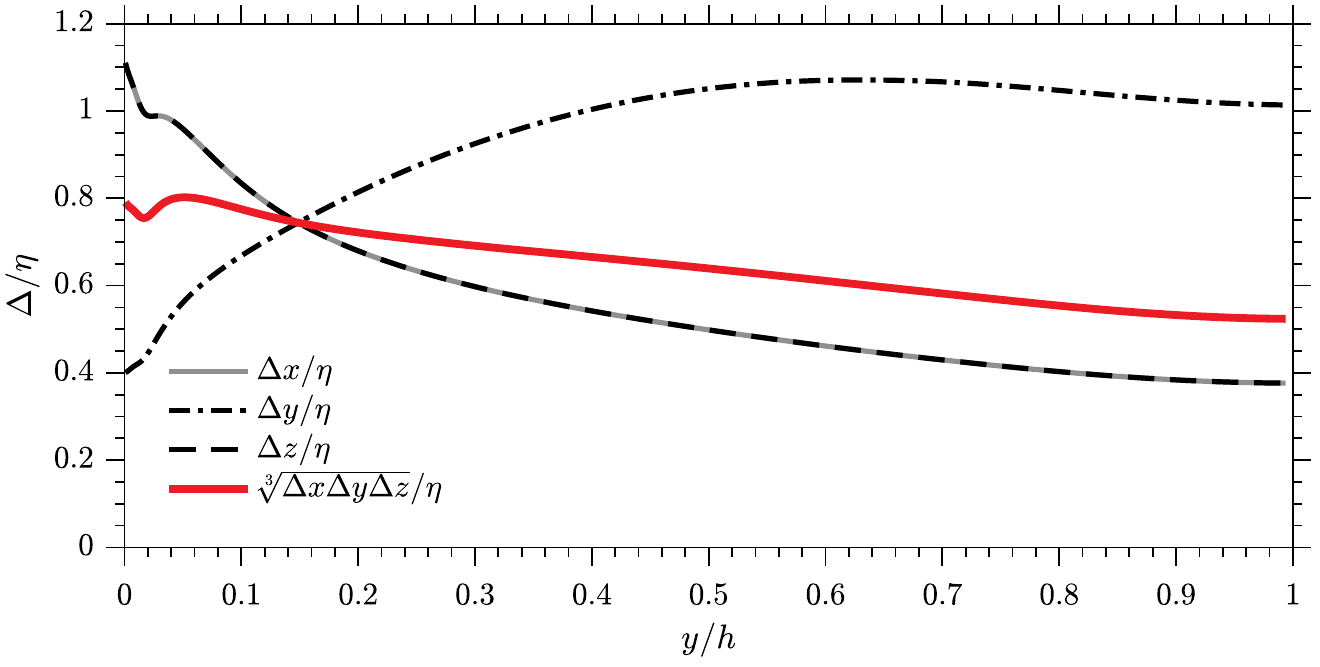}
\caption{Ratios between grid spacings and the Kolmogorov length scale across the wall-normal direction, for a plane channel, for the grid configuration G6.}
\label{fig:kolmogorov_scale}
\end{figure*}

As will transpire, the changes in drag relative to the plane channel are small, pushing the requirements for the spatial resolution to much more stringent levels than for regular DNS. 
Therefore, particular emphasis is placed on a few simulations performed at the highest tractable resolution within the resources available.
These key simulations correspond to the finest grid G6, at a bulk Reynolds number of $\Rey_b=6200$ ($\Rey_\tau \approx 360$).
Quantitative evidence for the level of refinement is given in \cref{fig:kolmogorov_scale} by scaling the grid dimensions by the Kolmogorov length scale in a plane-channel DNS for the mesh G6, showing that the ratio $\Delta/\eta$, where $\Delta = \sqrt[3]{\Delta x \Delta y \Delta z}$, remains lower than unity throughout the channel.

\begin{figure*}\centering
\vspace{-1cm}
\includegraphics[scale=1]{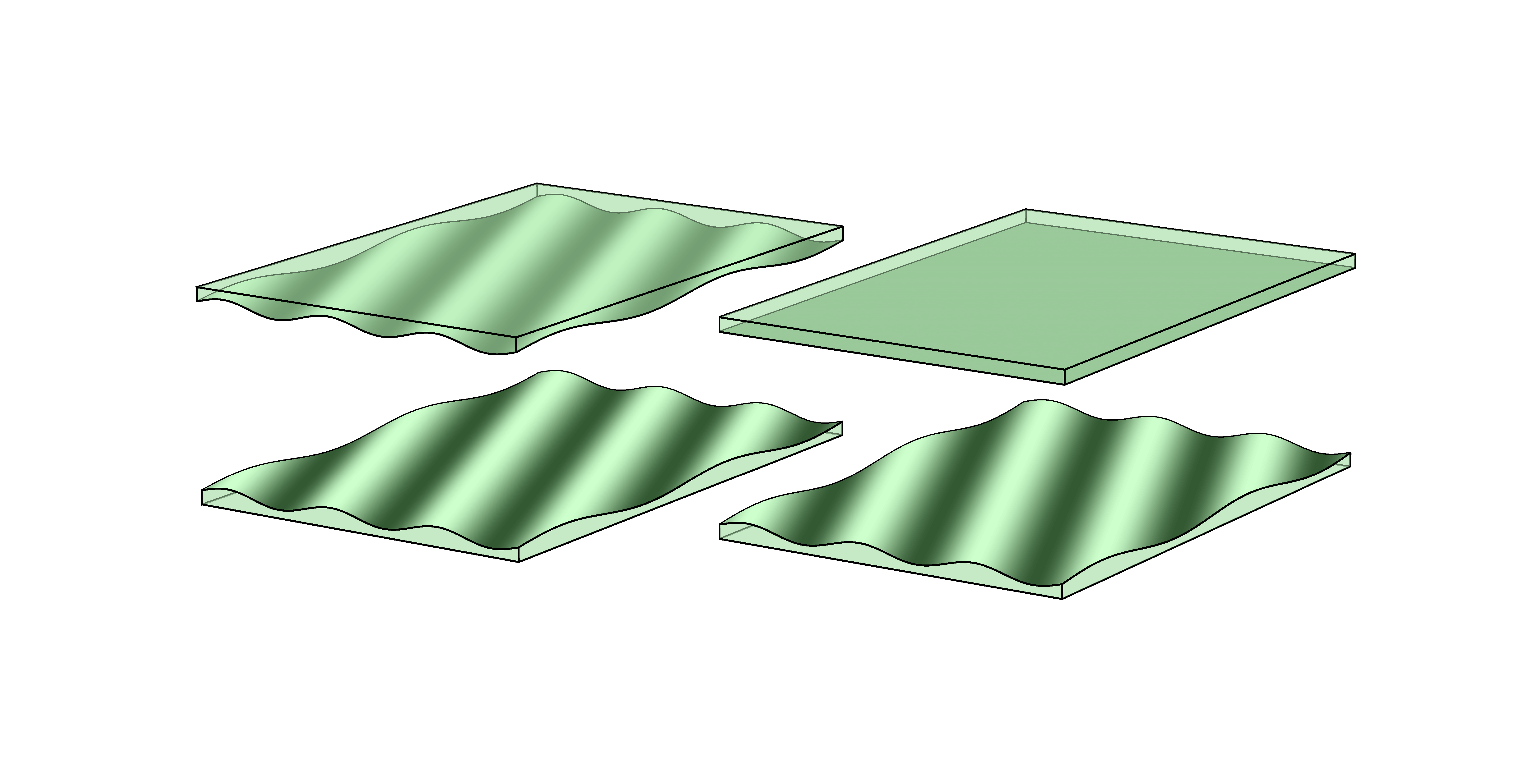}\\
\vspace{-7.02cm}
\includegraphics[scale=1]{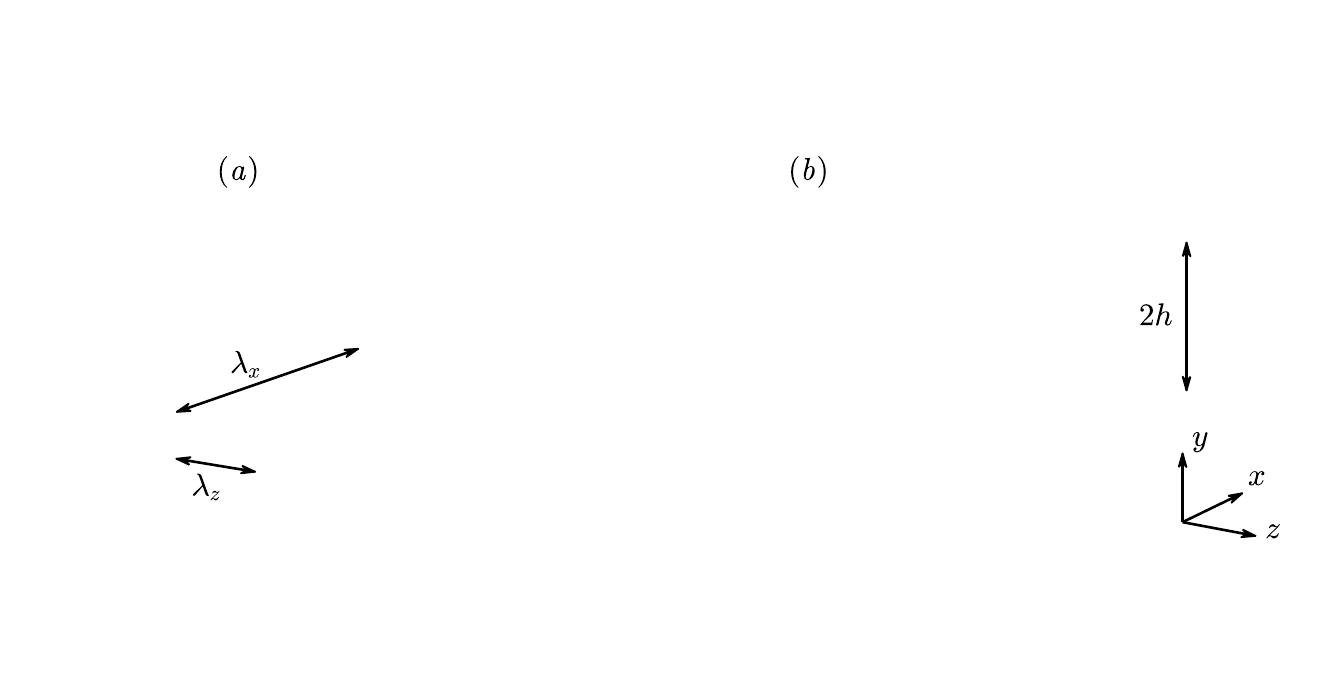}
\vspace{-1.6cm}
\caption{Sketch of the geometrical configurations: (\textit{a}) wavy-wavy channel~(w-w) and (\textit{b}) wavy-flat channel~(w-f). Here, the main flow direction is along the $x$-direction.}
\label{fig:wavywall}
\end{figure*}

The wavy mesh is generated by adding an increment to the wall-normal coordinate of each and every cell of the plane channel.
Two types of wavy geometry are considered herein: one with both walls wavy and the other with one wall flat, as shown in \cref{fig:wavywall}.
In the former, shown in \cref{fig:wavywall}\,\textit{a}, both walls are in phase, i.e.~yielding a constant passage height of $2h$ along the entire channel.
In the latter, shown in \cref{fig:wavywall}\,\textit{b}, the local height varies from $2h-A_w$ to $2h+A_w$, where $2h$ is the mean channel height, and $A_w$ the amplitude of the sinusoidal wall undulations.

\begin{table*}
\caption{Properties of the grid configurations, on which almost all the simulations presented are based. Viscous units are based on the target value of $\Rey_\tau$ for the plane channel. (Note that the angle of the flow may vary, e.g.~for an angle of $\pi/2$, the roles of $x$ and $z$ are reversed.)}
\label{tab:grids}
\begin{minipage}{0.9\textwidth}
\begin{ruledtabular}
\begin{tabular}{ccccccccccc}
 Grid label& $\Rey_\tau$ & $N_x$ & $N_y$ & $N_z$ & $\Delta x^+$ & $\Delta y^+$ & $\Delta z^+$ & $\Delta t^+$ & $L_x$ & $L_z$ \\ [3pt]
 \hline
 
G1 & 180  &  768 & 192 &  768 & 2.4 & 0.7 -- 2.9 & 2.4 & 0.04 & 10.2 & 10.2\\
G2 & 360  &  768 & 192 &  768 & 4.8 & 0.7 -- 7.3 & 4.8 & 0.05 & 10.2 & 10.2\\
G3 & 360  & 2208 & 192 & 2208 & 2.5 & 0.4 -- 8.5 & 2.5 & 0.02 & 15   & 15\\
G4 & 360  & 1104 & 192 & 1104 & 2.5 & 0.4 -- 8.5 & 2.5 & 0.02 & 7.5  & 7.5\\
G5 & 1000  & 1024 & 768 & 1152 & 7.1 & 0.5 -- 5   & 8.7 & 0.03 & 7.2  & 10\\
G6 & 360 &  1104 & 288 & 2208 & 1.7 & 0.6 -- 4.5 & 1.7 & 0.02 & 5.1 & 10.2\\
\end{tabular} 
\end{ruledtabular}
\end{minipage}
\end{table*}

\subsubsection{Computational implementation for skewed flow \label{sec:skewness}}
\begin{figure}\centering
\includegraphics[scale=.9]{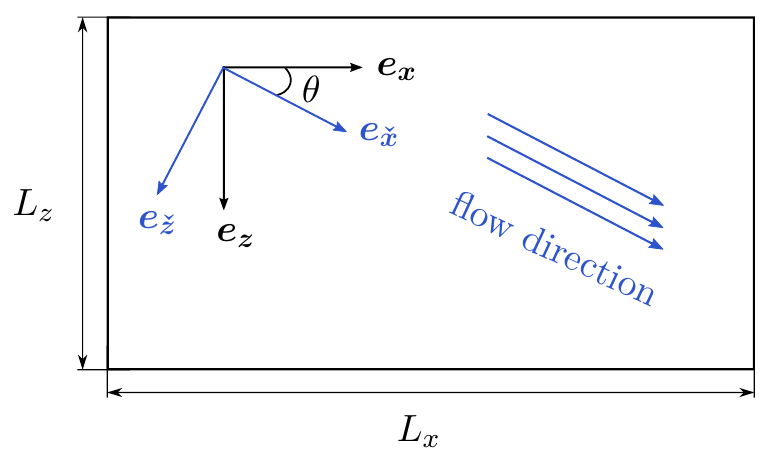}
\caption{Sketch of the configuration of the skewed-flow DNS, with flow-oriented coordinates $\check x$ and $\check z$, relative to domain coordinates $x$ and $z$.}
\label{fig:skewness}
\end{figure}

The skewed wavy channel can be simulated in two ways: the grid can be aligned with the wave or with the main flow direction.
Although the two are physically equivalent, keeping the wavy boundary aligned with the numerical box and skewing the flow at an angle to the grid allows greater flexibility.
Specifically, the main advantages of this approach are as follows:
\begin{enumerate}
\item if the crests are aligned with the $z$-direction, this direction becomes statistically homogeneous;
\item the post-processing is significantly eased; and 
\item this option allows continuous variations of the domain extent in the $z$-direction without affecting the periodicity boundary conditions applied to the $z$-direction boundaries.
\end{enumerate}
However, a disadvantage of this option is that it increases the problem size, since there is no longer an alignment between the $x$--$z$ coordinates and the streamwise and spanwise directions, respectively (cf.~\cref{fig:skewness}), necessitating both the domain sizes and resolution to be increased in the two wall-parallel ($x$--$z$) directions.
This is in contrast to the usual practice of increasing the spanwise resolution whilst decreasing the spanwise domain width.
Nevertheless, with the flow inclined at an angle to the grid, longer structures can be captured, as the flow traverses diagonally, thus mitigating the requirements for large domain sizes.
Additionally, the difference between the two orientation strategies is shown to be small in \cref{sec:gridcv}.

An approximately constant flow rate across the channel is maintained by iteratively updating the two orthogonal pressure gradients, $P_x$ and $P_z$, implemented as explicit body forces in the momentum equations, so that the bulk velocity is close to unity in the streamwise ($\check x$) direction and zero in the spanwise ($\check z$) direction.
The data shows that the target streamwise bulk velocity is satisfied within an error lower than $0.001\%$.
Throughout the discussion to follow, only the incremental part of the pressure is reported, relative to a pressure-reference value located at one of the corners of the computational box.

Quantities expressed in the frame of reference of the flow will be denoted with an overlaying inverted circumflex accent ($\check{\cdot}$), as in~\cref{fig:skewness}, although this notation will be omitted later when not needed.
Unless stated otherwise, all physical interpretations are given in the frame of reference aligned with the flow.

\subsection{Flow decomposition}

All statistical quantities can be averaged in the homogeneous direction parallel to the wave crests and troughs.
This groove-wise-averaging procedure is significantly eased by the choice made in~\cref{sec:skewness} of forcing the flow at an angle $\theta$ to the numerical grid.
Phase-averaging is also performed when multiple waves are included within the domain of solution.
Furthermore, in the case of a wavy channel with constant wall separation, both boundaries are statistically equivalent, which allows a doubling of the data included in the phase-averaging by shifting one of the walls by half a period and then taking advantage of the symmetry to average over both walls.
Thus, any time- and phase-averaged quantity $\overline{\boldsymbol{q}}$ only depends upon the phase location $x/\lambda$ and the wall-normal location $y$, reducing the dependence to $\overline{\boldsymbol{q}}({x}/{\lambda}, {y})$.

Depending on the objective of the analysis, two types of statistical decomposition of the mean turbulence properties can be considered.
The first decomposition is relevant to studying how the flow properties vary in phase:
\begin{equation}
\overline{\boldsymbol{q}} = \boldsymbol{Q} + \widetilde{\boldsymbol{q}}
\label{eq:triple1}
\end{equation}
where $\overline{\boldsymbol{q}}$ is any time- and phase-averaged quantity, $\boldsymbol{Q}(y) = \int_{0}^{1}\overline{\boldsymbol{q}}(x/\lambda,y+y_w)\, \dd (x/\lambda)$, ${y_w=A_w \sin(2\pi\,x/\lambda)}$ is the wall-normal location of the wall, and $\boldsymbol{\widetilde{q}}$ is the phase-varying part of the mean field.
The second approach lays emphasis on the action of the wavy boundaries relative to the plane-channel flow:
\begin{equation}
\overline{\boldsymbol{q}} = \boldsymbol{{Q}_0} + \widehat{\boldsymbol{q}}
\label{eq:triple2}
\end{equation}
where $\boldsymbol{{Q}_0}$ is the baseline plane-channel-flow value, and $\widehat{\boldsymbol{q}}$ is the difference to the plane-channel-flow solution.
These decompositions will be referred to as `phase-integrated' (${\boldsymbol{Q}}$), `phase-varying' ($\boldsymbol{\widetilde{q}}$), and `difference' ($\widehat{\boldsymbol{q}}$).

\subsection{Calculation of the drag contributions}\label{sec:DRtech}

A drag coefficient is defined for each contribution to the drag as
\begin{equation}
D_{\star} = \frac{\bar F_\star}{\frac12 \rho\, L_x\, L_z\, \| \boldsymbol{U_b}\| ^2},
\end{equation}
where $\star$ identifies the contribution (e.g. `$f$' for friction), $\bar F_\star$ is the mean force exerted on the walls opposing the flow direction, and $\boldsymbol{U_b}$ the bulk velocity.

As mentioned in \cref{sec:skewness}, in all the cases presented, the flow is driven at approximately constant flow rate, so that $\|\boldsymbol{U_b}\| \approx 1$, via the imposition of a spatially-constant (vectorial) pressure gradient in order to balance the total drag force.
Given a unit bulk velocity, the correct Reynolds number is set by prescribing the appropriate viscosity.
The resulting pressure force driving the flow is:
\begin{equation*}
\bar F_{tot} = -P_{\check{x}}\, L_x \, 2h \, L_z,
\end{equation*}
where $P_{\check{x}}$ is the projection of the pressure gradient onto the flow direction: ${P_{\check{x}} = P_x\, \cos \theta + P_z \, \sin \theta}$.

For the wavy channel, the total drag force is composed of two contributions: friction and pressure drag.
The friction and pressure forces were integrated on the wavy surface and projected onto the flow direction, yielding the drag coefficients $D_f$ and $D_p$, respectively.

Since only pressure and friction forces act on the walls, the total drag is:
\begin{equation}
D_{tot} = D_p + D_f.
\label{eq:bal}
\end{equation}

\section{Simulations}

\subsection{Overview of simulations}
Simulations were performed at $\Rey_\tau \approx 360$ for various configurations, grid resolutions, and domain sizes.
This Reynolds number was chosen so that the cost of the simulation remained tractable, and that the ratio between the height of the wave and the channel height $A_w/h$ was kept relatively modest.
The corresponding parameters are given in \cref{tab:drag}, along with the drag levels calculated, as explained in \cref{sec:DRtech}.
The net drag variation is evaluated in two ways: from the imposed pressure gradient and from the sum of the surface-integrated pressure force and friction-drag force.
As expressed by \cref{eq:bal}, the two should be identical. However, they differ very slightly in the simulations, and this is expressed by the column `imbal.' in \cref{tab:drag}.
In all simulations, the imbalance of the forces is regarded as negligible.
By way of contrast with previous DNS studies reporting this value, the force imbalance in \cite{Wang2006} was of about 3--4\%.

{\setstretch{1.0}
\begin{table*}
  \caption{Configurations simulated, all at ${\Rey_\tau \approx 360}$. Each simulation is designated by a label consisting of a set of letters plus identifiers, meant to convey as much information as possible in a compact manner, and identifying each calculation uniquely. `G' stands for `Grid' and refers to the grid configurations detailed in \cref{tab:grids}, `W' stands for `Wavy', `P' for `Plane', and `A' for `Amplitude'. The figure following the letter `A' identifies a particular value of the wave slope $A_w/\lambda$.  The suffix `f' indicates the presence of a flat upper wall instead of two in-phase wavy walls, and `bis' reflects the fact that the wavelength of `G2W1bis' is equal to that of `G2W1', but the flow angle $\theta$ is different. TDR, PD and FDR respectively stand for Total-Drag Reduction, Pressure Drag and Friction-Drag Reduction.}
    \begin{ruledtabular}
    \begin{tabular}{lrccccrrrrrrr}
\multicolumn{1}{c}{\textbf{Simulation}} & \multicolumn{4}{c}{\textbf{Flow}} &  \multicolumn{4}{c}{\textbf{Drag coefficients} } & \multicolumn{3}{c}{\textbf{Relative}}\\
\multicolumn{1}{c}{\textbf{label}} & \multicolumn{4}{c}{\textbf{configuration}} &  \multicolumn{4}{c}{($\times 10^6$)} & \multicolumn{3}{c}{\textbf{drag variation}}\\ [5pt]
\hline
     & \multicolumn{1}{c}{$A_w^+$} & \multicolumn{1}{c}{$\theta (^\circ)$} & \multicolumn{1}{c}{$\lambda^+$} & \multicolumn{1}{c}{$\lambda_x^+$} &  \multicolumn{1}{c}{$D_{tot}$} & \multicolumn{1}{c}{$D_{p}$} & \multicolumn{1}{c}{$D_f$} & \multicolumn{1}{c}{imbal.} & \multicolumn{1}{c}{TDR} & \multicolumn{1}{c}{PD} & \multicolumn{1}{c}{FDR}\\ [3pt]
\hline
\multicolumn{12}{c}{Key simulations}\\
    ~G6P1  & 0     & 70    & --     & --      & 6679  & 0     & 6677  & -0.03\% &  &  &\\
    ~G6W1A1 & 11    & 70    & 918   & 2684    & 6659  & 30    & 6628  & -0.02\% & ~0.3\% & 0.45\% & 0.74\% \\
    ~G6W1A2 & 18    & 70    & 918   & 2684   & 6633  & 87    & 6544  & -0.02\% & ~0.7\% & 1.30\% & 1.99\% \\
    ~G6W1A3 & 22    & 70    & 918   & 2684    & 6639  & 130   & 6507  & -0.03\% & ~0.6\% & 1.95\% & 2.55\% \\
    ~G6W1A4 & 32    & 70    & 918   & 2684    & 6728  & 332   & 6394  & -0.03\% & -0.7\% & 4.97\% & 4.24\% \\
    ~G6W2A1 & 7     & 70    & 612   & 1789    & 6671  & 32    & 6637  & -0.02\% & 0.1\% & 0.48\% & 0.60\% \\
    ~G6W2A3 & 14    & 70    & 612   & 1789    & 6676  & 138   & 6539  & 0.00\% & 0.0\% & 2.07\% & 2.07\% \\
    ~G6W2A4 & 22    & 70    & 612   & 1789    & 6780  & 329   & 6450  & -0.01\% & -1.5\% & 4.92\% & 3.40\% \\ [5pt]
\multicolumn{12}{c}{Other simulations}\\
    ~G3P1  & 0     & ~0     & --    & --      & 6642  & 0     & 6639  & -0.04\% &  &  &  \\
    ~G3P2  & 0     & 45    & --    & --      & 6690  & 0     & 6688  & -0.03\% &  &  & \\
    ~G3W1  & 18    & 70    & 918   & 2684    & 6622  & 87    & 6533  & -0.04\% & 1.1\% & 1.29\% & 2.41\%\\
    ~G4P1  & 0     & ~0     & --  & --    & 6640  & 0     & 6637  & -0.04\% &  &  & \\
    ~G4P2  & 0     & 45    & --  & --    & 6694  & 0     & 6691  & -0.04\% &  &  & \\
    ~G2P1  & 0     & 52    & --   & --   & 6805  & 0     & 6809  & -0.01\% &  &  & \\
    ~G2P2  & 0     & 70    & --   & --   & 6755  & 0     & 6755  & 0.01\% &  &  & \\
    ~G2W1  & 18    & 52    & 918   & 1491   & 7021  & 544   & 6479  & 0.02\% & -3.0\% & 7.99\% & 4.85\% \\
    ~G2W1f & 18    & 52    & 918   & 1491   & 6911  & 271   & 6641  & 0.01\% & -1.5\% & 3.98\% & 2.47\% \\
    ~G2W1bis & 18    & 70    & 918   & 2684   & 6648  & 87    & 6560  & 0.00\% & 1.6\% & 1.29\% & 2.89\% \\
    ~G2W2  & 14    & 70    & 612   & 1789   & 6640  & 138   & 6503  & 0.01\% & 1.7\% & 2.04\% & 3.73\% \\ 

    \end{tabular}%
    \end{ruledtabular}
  \label{tab:simulations}%
  \label{tab:drag}
\end{table*}%
}

\subsection{Overall physical characteristics}
\begin{figure*}
\includegraphics[scale=\scalefig]{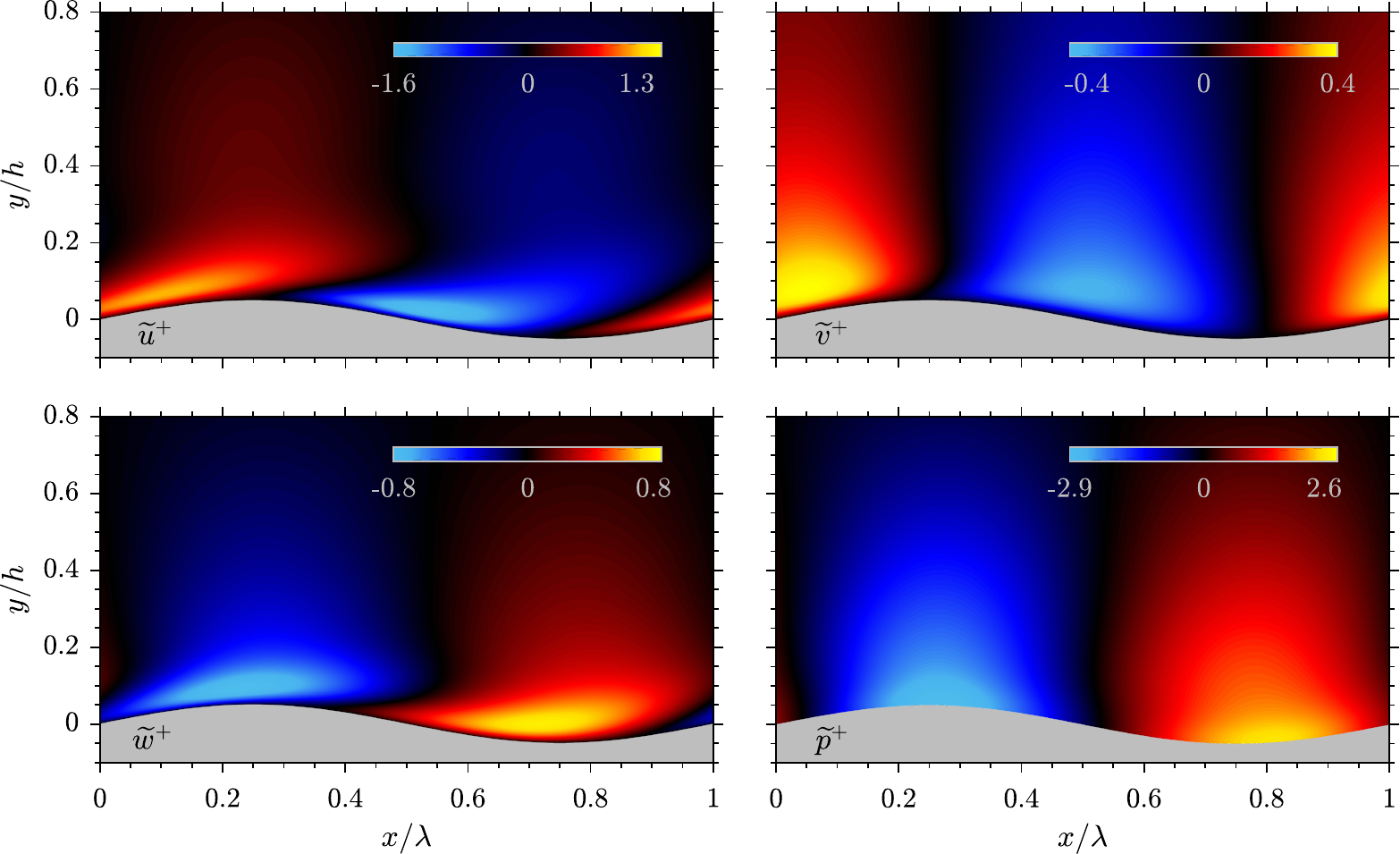}
\caption{Contours of the phase-varying velocity and pressure fields for G6W1A2 ($A_w^+=18$, $\theta=70^\circ$, $\lambda^+=918$).}
\label{fig:velp}
\end{figure*}

\begin{figure*}\centering
\includegraphics[scale=\scalefig]{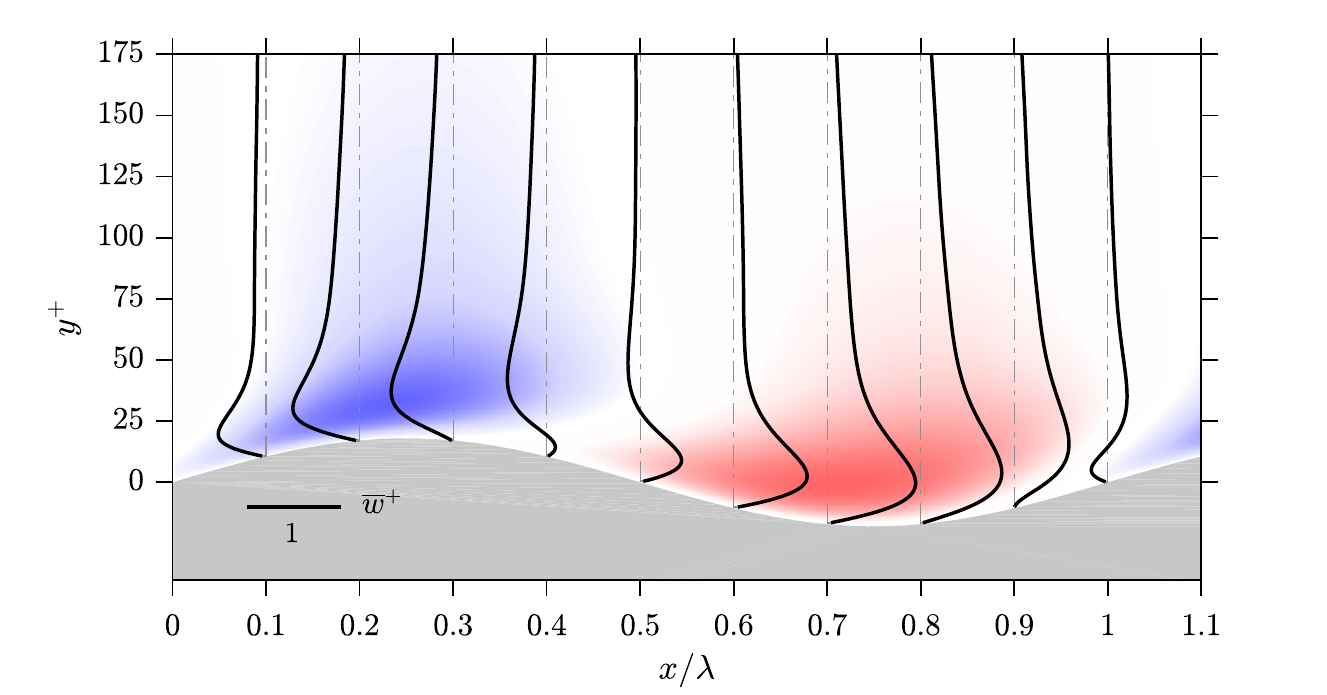}
\caption{Mean spanwise motion along the wavy wall for G6W1A2 ($A_w^+=18$, $\theta=70^\circ$, $\lambda^+=918$). Black lines: spanwise-velocity profiles at evenly-spaced phase locations, background: contours of the spanwise velocity.}
\label{fig:spawisemo}
\end{figure*}

\begin{figure*}
\includegraphics[width=.87\textwidth]{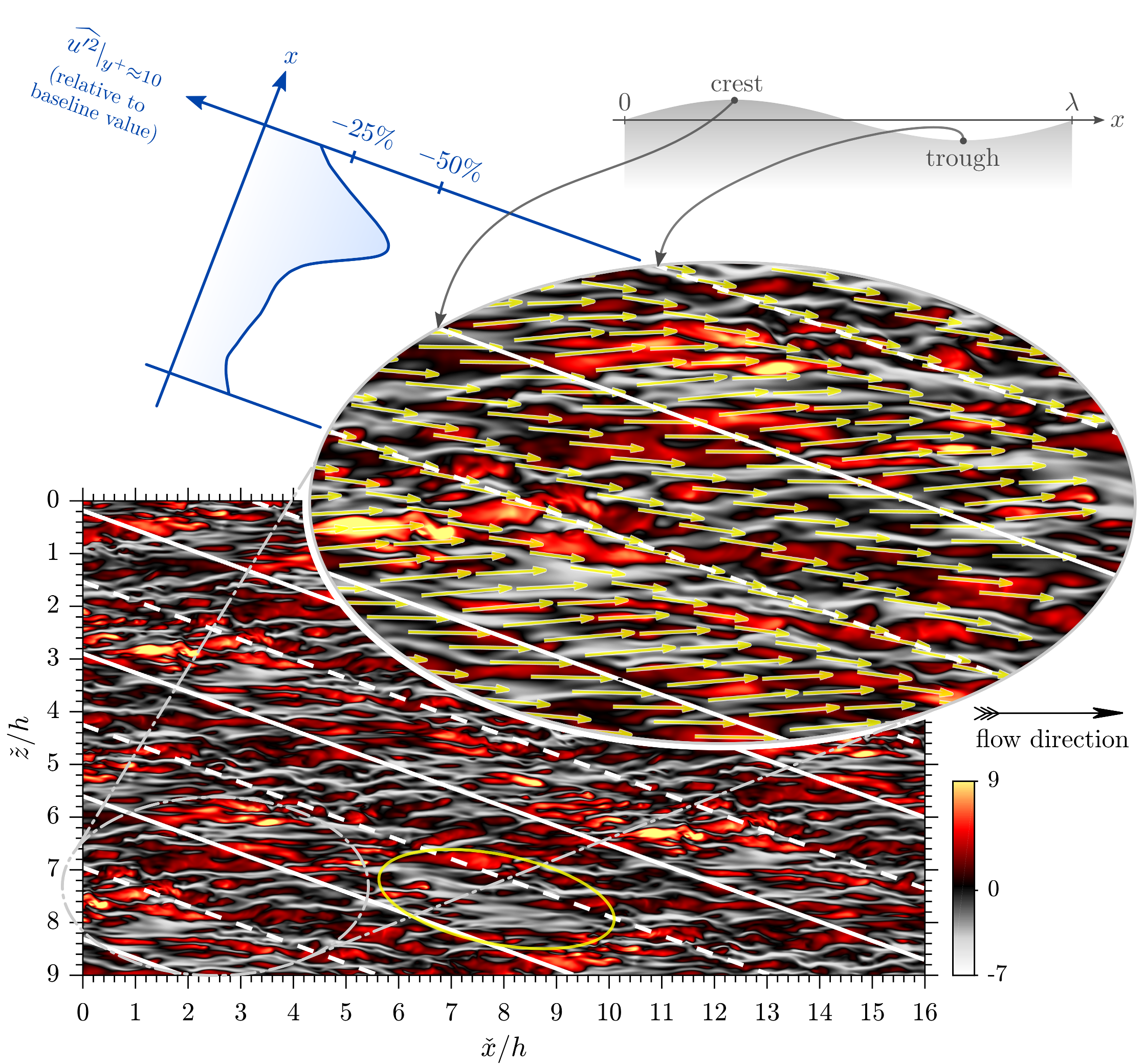}
\caption{Turbulent fluctuations of the streamwise velocity $u'^+$ for G6W1A4 (${A_w^+=32}$, ${\theta=70^\circ}$, ${\lambda^+=918}$), normalised by the local mean friction velocity at the particular phase location $\sqrt{\nu (\partial \overline{u}/\partial y)|_w}$ in a wavy horizontal slice located at constant $y/h$, around $y^+\approx 10$. Straight white lines: location of the crests, dashed straight white lines: location of the troughs, yellow arrows: mean shear-strain field, yellow circle: region of strong streak weakening. The periodic boundary conditions have been exploited to increase the visualisation area.}
\label{fig:snap}
\end{figure*}

As the flow travels past the skewed wavy wall, it accelerates on the windward side and then decelerates on the leeward side, a behaviour linked to the pressure being a minimum above the crest and a maximum in the trough region, as shown in \cref{fig:velp}.
Because the wave is at an angle to the main flow direction, this pressure variation in phase gives rise to a pressure gradient both in the streamwise and spanwise direction.
The latter gradient generates a spanwise motion, shown in \cref{fig:spawisemo}, which is asymmetric in phase and penetrates quite far into the boundary layer above $y^+\approx 100$.
However, as previously mentioned, it is not the velocity but the shear strain which is important with respect to the emulation of the Stokes layer, and which dictates the orientation of the near-wall streaks.
In \cite{Touber2012}, this orientation was observed to vary in time, with strong reduction in turbulence during the reorientation phase.
For the wavy wall, the phase-modulation of the shear strain occurs in space, and also results in a reorientation following the shear-strain field, as well as in a weakening of the streaks, as shown in \cref{fig:snap} at a constant distance from the wall around $y^+\approx 10$. 
However, the effect is far less pronounced than observed in \cite{Touber2012} because the forcing amplitude is much smaller in the present case.

\subsection{Influence of the upper wall}\label{sec:flat}
Previous studies on wavy channels often had only a single wavy wall, or featured varicose wall undulations \citep{Cherukat1998,Henn1999,Yoon2009,Nakanishi2012,Mamori2012,Luchini2016}.
A distinct feature of the present configuration is that the passage height is constant at any phase location.
Thus, comparing various wall configurations is interesting in order to ensure that the distance between the two walls is large enough, so that the upper wavy wall does not influence significantly the flow above the lower one.
This is addressed by comparing simulations for channels with one or both walls wavy.
The relevant entries in \cref{tab:drag,tab:drag_flat} are G2W1 and G2W1f, and the corresponding plane-channel baseline G2P1.
The results in~\cref{tab:drag_flat} show that the friction on the lower (wavy) boundary of G2W1f has a lower value than for G2W1, whilst the friction on the upper (flat) surface of G2W1f is slightly increased with respect to the baseline drag.
The latter is a consequence of the control strategy of keeping the flow rate constant.
The drag coefficients reported in~\cref{tab:drag_flat} show that the drag relative to the baseline level (in this case, drag increase) is about half of that of the case with both walls wavy: the drag increase per wavy wall in G2W1 is 1.58\%, compared to 1.54\% for the wavy-flat case G2W1f.
This supports the assertion that the two configurations are close to each other in terms of the processes effective at each wavy wall separately.

\begin{table}
    \caption{Drag coefficients at both upper and lower walls, comparing a wavy-wavy channel (G2W1) to its wavy-flat counterpart (G2W1f).}
\label{tab:drag_flat}
  \begin{ruledtabular}
    \begin{tabular}{lccccc}
         \multicolumn{1}{c}{Simulation} & \multicolumn{2}{c}{$D_f$ ($\times 10^6$)} & \multicolumn{2}{c}{$D_p$ ($\times 10^6$)} & DR\\
         \multicolumn{1}{c}{label} & \multicolumn{1}{c}{lower} & \multicolumn{1}{c}{upper} & \multicolumn{1}{c}{lower} & \multicolumn{1}{c}{upper} & (total)\\
\hline
    ~G2P1  & 3405  & 3402  & {--} & -- & -- \\
    ~G2W1  & 3242  & 3237  & 272   & {271} & -3.16\%\\
    ~G2W1f & 3220  & 3421  & 271   & -- & -1.54\%\\
    \end{tabular}%
  \end{ruledtabular}
\end{table}

\subsection{Net drag reduction and grid convergence}\label{sec:gridcv}
In \cref{tab:drag}, TDR represents the drag relative to that of the plane channel. Physically, the latter is unique regardless of the angle between the flow an the mesh.
However, this is not so computationally, owing to variations in solution domain and numerical errors, including the finite time of averaging, grid resolution, and the finite size and orientation of the periodic domain which does not allow very long structures to be captured.
As the angle between the main flow direction and the grid varies, the length of the longest structure allowed to exist within the periodic boundaries changes.
At the same time, the grid resolution is also altered in the flow-oriented directions.
This results, in a plane channel, in a slight dependence of the drag on the flow direction.
By way of example of the influence of the domain size on the drag, for a flow aligned with the mesh, \citet{Ricco2004} observed, at $\Rey_\tau = 200$, that changing the streamwise extent from $4\pi h$ to $21h$ whilst keeping roughly the same spatial resolution, resulted in a drag increase of 0.6\%.
This difference in drag is already large enough to be of the same order of magnitude as the variations of the total drag level in the cases considered herein.
Therefore, a careful assessment of key computational parameters is required, as detailed below.

First, the influence of the domain size is considered for plane-channel flow.
This was investigated by means of highly-resolved simulations, with constant resolution, but varying domain sizes.
For this purpose, grids G3 and G4 are chosen to have the same spatial resolution $\Delta x^+ = \Delta z^+ = 2.5$ and $\Delta y^+$ ranging from 0.4 at the wall up to 8.5 in the centre, but the domain size of G4 is one half of that of G3 in the $x$-$z$ directions (cf.~\cref{tab:grids}).
The largest domain considered is $L_x = L_z = 15h$, which represents about 5400 wall units, i.e.~longer than the commonly chosen value of $8\pi h^+=O(4500)$ at ${\Rey_\tau = O(180)}$.
\Cref{tab:drag} shows that the relative difference in drag between the larger domain (grid G3) and the smaller (grid G4) is lower than 0.05\%, both at an angle of $0^\circ$ (G3P1 and G4P1) and $45^\circ$ (G3P2 and G4P2).

Next, the impact of resolution is quantified, still for the case of a plane channel.
To this end, simulations were run with the domain size kept constant, as in G3, but with a mesh coarser by a factor of 2 in the wall-parallel and wall-normal directions independently.
Then, the drag levels for $\theta = 0^\circ$ and $45^\circ$ configurations were compared.
As expected for a consistent discretisation, the difference between the two physically-equivalent configurations ($\theta = 0^\circ, 45^\circ$) decreases as the resolution is increased.
However, this difference remains significant at about 0.7\% for grid G3, despite the mesh being fine with respect to usual DNS standards.
An important observation is that increasing the number of cells in the wall-normal direction has little impact on the total drag, whereas increasing the resolution in the wall-parallel directions reduces significantly the difference in friction between $\theta= 0^\circ$ and $\theta = 45^\circ$.
Drag differences with respect to the angle of the flow were observed to vary monotonically, the minimum drag coefficient being found at $\theta = 0^\circ$, and the maximum at $\theta = 45^\circ$.

A possible approach to reducing the error in the predicted drag-reduction level is to evaluate it by reference to a plane-channel flow simulation at exactly the same spatial resolution, domain size, and flow angle.
This is the approach preferred here, in light of the work of \citet{Gatti2013,Gatti2016}, who observed some cancellation of the systematic bias associated with the domain size.
Thus, for example, the baseline drag for G2W1 is G2P1 (both at $\theta = 52^\circ$), whereas for G2W1bis ($\theta = 70^\circ$), the baseline is taken to be G2P2 (also at $\theta = 70^\circ$). 
However, even this elaborate approach may not suffice, in the face of the small drag-reduction margin, to remove uncertainties associated with the numerical aspects that contribute to the total error.
A factor that may also be influential is the distortion of the cells fitted to the wavy boundary.
As shown in~\cref{tab:draggridconv}, changes in the total drag from grid G2 to the finest grid G6 are not independent of the wave geometry.

\begin{figure*}\centering
\includegraphics[scale=\scalefig]{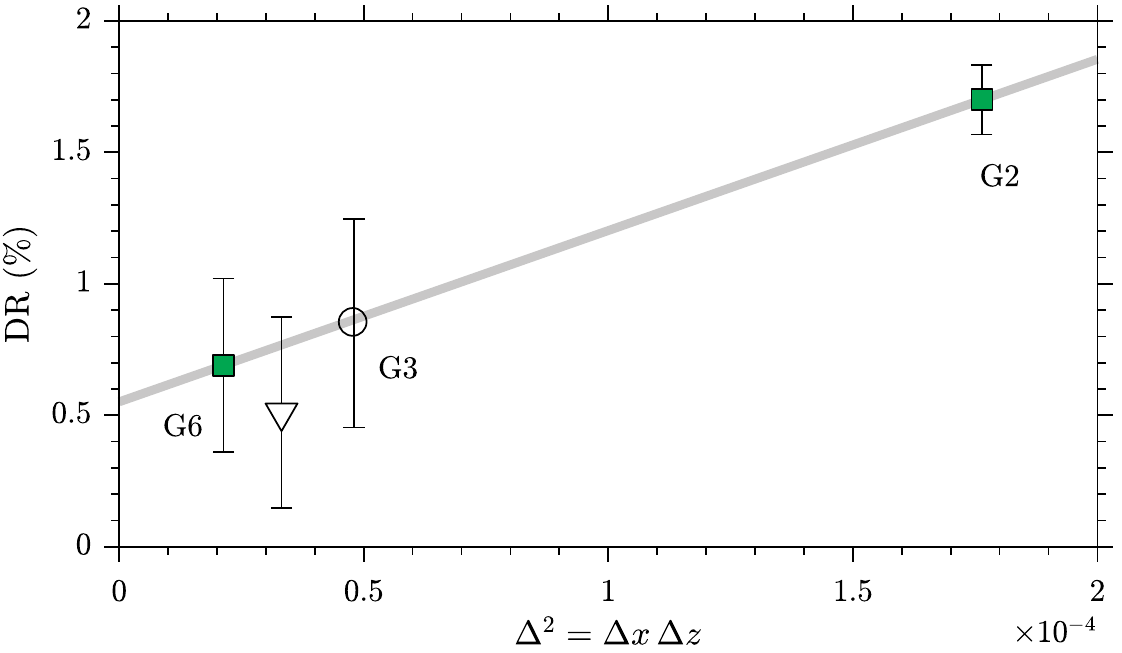}
\caption{Total-drag reduction for $A_w^+=18$, $\theta = 70^\circ$, and $\lambda^+=918$, for grids G2, G3 and G6. Inverted triangle $\triangledown$: supplementary run with the wave at an angle and the flow aligned with the grid. The baseline drag level is taken at the same angle, except for G3 where the baseline drag is found by interpolation between $\theta=0^\circ$ and $\theta=90^\circ$ cases. Error bars: 90\% confidence interval of the time-averaging error calculated using the method of batch means and batch correlations by \citet{Luchini2017}, and assuming that the baseline and controlled drag levels are independent variables.}
\label{fig:grid_cv}
\end{figure*}

The variation of the total-drag reduction with grid refinement was studied in greater detail for the most promising case, namely ${(A_w^+, \theta, \lambda^+) = (18,70^\circ,918)}$.
Since the drag was found to be mainly sensitive to the wall-parallel resolutions, only $\Delta x$ and $\Delta z$ are used as indicators of the grid refinement.
The main outcome of the study, shown in~\cref{fig:grid_cv}, is that the total-drag reduction predicted decreases as the mesh is refined, with a quadratic dependence on the wall-parallel mesh spacing.
The drag-reduction value appears to tend, asymptotically, to a positive value of 0.5\%.
Additionally, the ratios of the friction- and pressure-drag coefficients ($D_f$ and $D_p$ respectively), relative to the total $D_{tot}$, remain constant for grids G2 and G6: thus, the value of $D_p$ is 1.314\% that of $D_{tot}$ for the coarsest mesh G2W1A2, compared with 1.315\% for the finest mesh G6W1A2.
Therefore, whilst the share between the pressure and friction drag is grid-independent, the absolute value of the total drag is a very sensitive quantity that requires substantial computational efforts to be accurately predicted.

\begin{table}
    \caption{Grid refinement study for wavy calculations, all at $\theta = 70^\circ$.  The coarsest grid is G2, and the finest is G6. Abbreviations are consistent with those of \cref{tab:drag}.}
  \label{tab:draggridconv}%
  
	\begin{ruledtabular}
    \begin{tabular}{lccccc}
        \multicolumn{1}{c}{Simulation label}  & $\Delta x^+$ & $\Delta z^+$ & FDR   & PD    & TDR \\
\hline
    ~G2W1bis & 4.8 & 4.8 & 3.0\% & 1.3\% & 1.7\% \\
    ~G3W1  & 2.5  & 2.5 & 2.4\% & 1.3\% & 1.1\% \\
    ~G6W1A2 & 1.7 & 1.7 & 2.0\% & 1.3\% & 0.7\% \\
\hline    
    ~G2W2  & 4.8 & 4.8 & 3.8\% & 2.0\% & 1.8\% \\
    ~G6W2A3 & 1.7 & 1.7  & 2.1\% & 2.1\% & 0.0\% \\
    \end{tabular}%
    \end{ruledtabular}
\end{table}%

One additional simulation, physically equivalent to G6W1A2, was undertaken, although at a slightly coarser resolution, in a computational box aligned with the main flow direction, and the wavy wall at an angle to the grid.
Two wavelengths were included in the streamwise and spanwise directions.
The resulting point, shown in~\cref{fig:grid_cv}, is in line with the simulations of the skewed configuration.

\subsection{Friction-drag reduction and pressure-drag increase}\label{sec:FDR}

\begin{figure*}
\begin{center}
{\hspace{0pt}\includegraphics[scale=\scalefig]{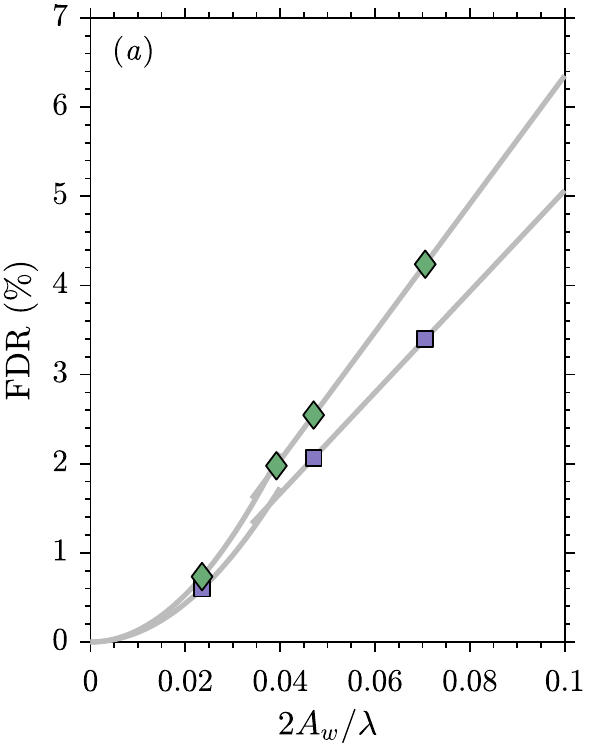}}\quad%
{\hspace{10pt}\includegraphics[scale=\scalefig]{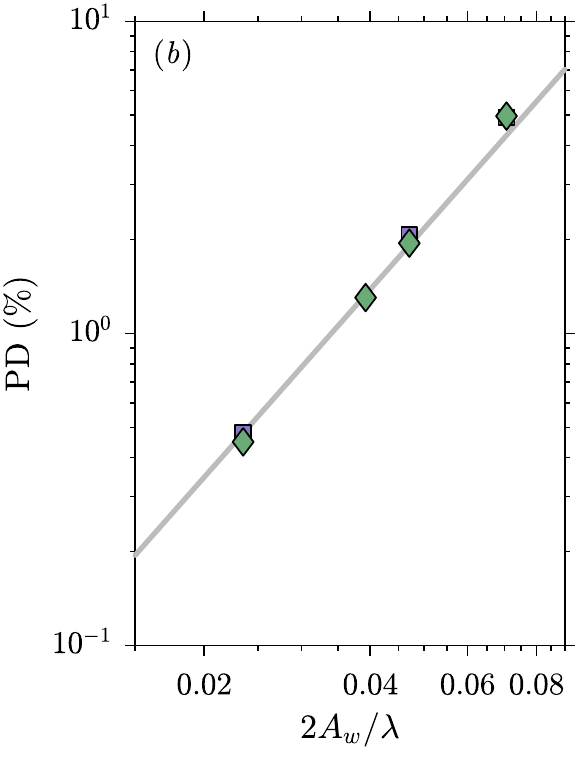}}%
\end{center}
\caption{(\textit{a}) Levels of friction-drag reduction for simulations G6W*A*.  $\blacklozenge$ correspond to G6W1A* $({\lambda^+ = 918})$, and $\blacksquare$ to G6W2A* $({\lambda^+ = 612})$.  Continuous lines represent quadratic and linear behaviours. The quadratic curve is interpolated from the value at ${2A_w/\lambda=0.024}$ with zero values for the FDR and its derivative at ${A_w/\lambda = 0}$. The linear curve is interpolated from the two largest wave slopes. (\textit{b}) Levels of pressure drag for simulations G6W*A*.  Symbols: same as~(\textit{a}), line: quadratic behaviour.}
\label{fig:DR}
\end{figure*}

In \cite{Viotti2009}, the reduction in skin friction was observed to increase linearly with forcing amplitudes up to $A^+_\text{SSL} \approx 5$, and then to increase at a slower rate.
Such an observation is not consistent with the expected symmetry of the problem around $A^+_\text{SSL}=0$, which would imply a zero value of the first derivative of the drag reduction as a function of the amplitude.
For the wavy wall, the dependence of the friction-drag reduction (FDR) on the wave slope, shown in~\cref{fig:DR}\,(\textit{a}), is compatible with a symmetry condition, thus indicating that this is a local effect taking place below $2A_w/\lambda \approx 0.04$: for very small actuation amplitudes, the FDR does seem to exhibit a quadratic behaviour, which then becomes close to linear.
The amplitude of the spanwise shear strain at $2 A_w/\lambda \approx 0.04$ corresponds to that of a SSL with a forcing amplitude of about $A^+_\text{SSL} \approx 1.1$, thereby corroborating the observations of \cite{Viotti2009}, since the smallest amplitude they considered was $A^+_\text{SSL}=1$, which occurs at the intersection between the quadratic and linear behaviour for the present key simulations.

The decrease in $\lambda_x$ from G6W1* (${\lambda^+ = 918}$, ${\lambda_x^+ \approx 2700}$) to G6W2* (${\lambda^+=612}$, ${\lambda_x^+ \approx 1800}$) results in a reduced effectiveness of the wavy wall at the same wave slope, i.e. the same equivalent forcing amplitude $A_\text{SSL}^+$.
This trend contrasts with the drag-reduction trend of the SSL, which features an optimum around ${\lambda_x^+ \approx 1250}$.
Therefore, there exists some unfavourable mechanism limiting the drag reduction achievable by the wavy wall.
Such considerations will be discussed in \cref{sec:ssldiff}.

\section{Emulation of a spatial Stokes layer}
\subsection{Shear strain}\label{sec:shearcmp}
The present approach of using a skewed wavy wall is based on the assumption that similar longitudinal patterns of the transverse shear strain, whether created by the SSL or the wavy wall, will lead to the same effects on turbulence and hence lead to some turbulent-drag reduction.
In the case of a temporal Stokes layer, \citet{Touber2012} showed that the wall oscillations led to a bimodal partial decay, reformation and reorientation of the streaks, a behaviour dictated by the unsteady Stokes strain in the buffer layer.
If this mechanism is indeed intimately linked to the friction-drag reduction, the relevant forcing quantity is the resulting shear strain, rather than the forcing velocity itself.
It is this key element that allows for a passive surface, with no slip at the solid wall, to emulate the actuation by in-plane wall oscillations, through the action of a transverse pressure gradient that generates an equivalent shear-strain field.

In contrast to the Stokes layer, the wavy wall also induces wall-normal forcing that contributes to the shear strain via ${\partial \overline{v}/\partial z}$, but this additional effect was observed to be small compared to ${\partial \overline{w}/\partial y}$, especially in the near-wall region where the shear is maximum.
It follows that there is no significant parasitic effect of the wall-normal velocity on the spanwise forcing, justifying a direct comparison of ${\partial \overline{w}/\partial y}$ between wavy-wall and Stokes-layer configurations.

A slight difference between the SSL simulation presented in this section, and those of \cite{Viotti2009}, has been introduced deliberately, in order to maximise the correspondence between the SSL and the wavy-wavy channel configuration.
This difference is that the wall forcing on the upper wall is shifted by half a period relative to the lower wall.
However, this change does not result in noticeable differences, since the thickness of the Stokes layer is much smaller than the channel half-height.
The reference SSL considered is for a forcing amplitude of $A^+_\text{SSL}=2$ (based on unactuated friction velocity) and a wavelength close to the optimum at this forcing amplitude, $\lambda_x^+ \approx 1250$, subject to the assumption that the Reynolds-number change from ${\Rey_\tau = 200}$ in~\cite{Viotti2009}, to the present value of ${\Rey_\tau \approx 360}$ does not have a significant impact on the optimal wavelength.
The simulation was run on a domain ${L_x^+\approx 2500}$, ${L_z^+\approx 1100}$, with a grid resolution $\Delta x^+ = 9.8$, $\Delta z^+ = 5.8$, and $0.7 < \Delta y < 7.3$.
Such a resolution may not be sufficient for an accurate comparison of the drag levels, but is acceptable for the comparison of the shear-strain profiles.

\begin{figure*}\centering
\includegraphics[scale=\scalefig]{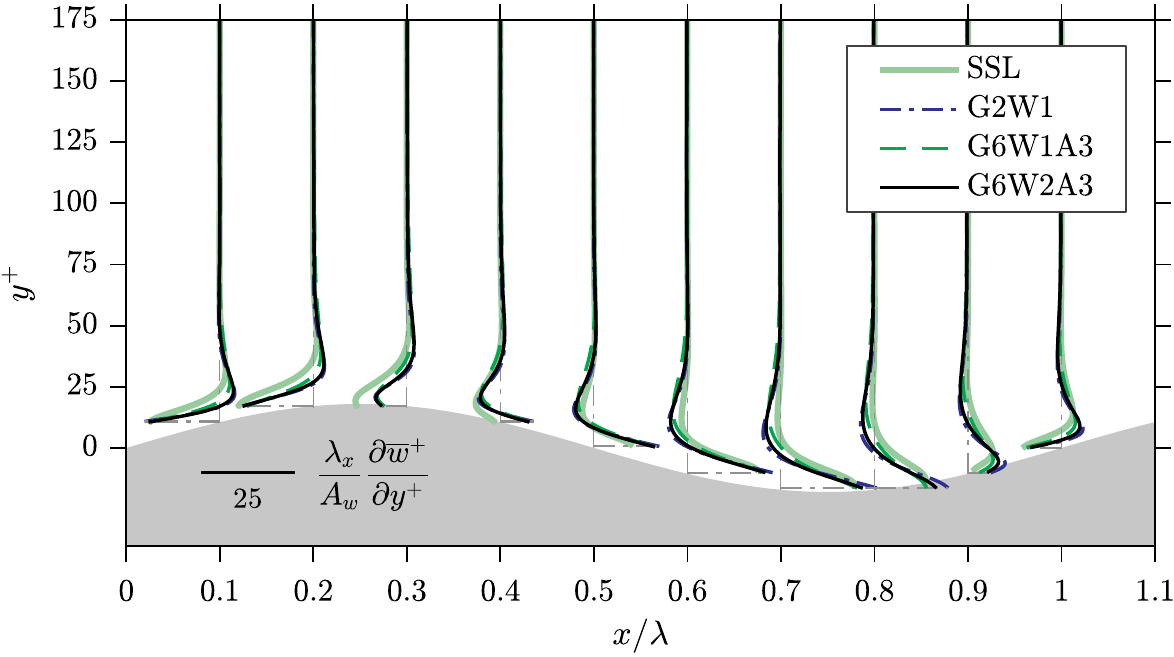}
\caption{Comparison of $\partial \overline{w}^+/\partial y^+$ scaled by the streamwise-projected wave slope $A_w/\lambda_x$ for various flow configurations. The SSL flow is for a wavelenth $\lambda_x\approx 1250$ and forcing amplitude $A_\text{SSL}^+ = 2$ based on the unactuated friction velocity. The value of $\partial w^+/\partial y^+$ is based on the actual friction velocity, and the scaling factor $A_w/\lambda_x$ based on the parameters of G2W1. In all cases, the profiles were shifted in the wall-normal direction in order to match the wave height of G2W1.}
\label{fig:cmp_shear}
\end{figure*}

Results for some wavy channels are shown in \cref{fig:cmp_shear}.
The strain profiles demonstrate that the wavy wall emulates reasonably well the shear layer of the SSL.
This observation leads to the expectation that the reduction in turbulent skin friction achieved by the wavy wall would be similar, and arises from the same physical mechanism as in the SSL.
In addition, the amplitude of the phase-wise variations in the shear-strain -- and hence the corresponding SSL forcing amplitude $A_\text{SSL}^+$-- appears to be mostly dictated by the streamwise wave slope $A_w/\lambda_x$, as suggested by rescaling the amplitude of the transverse shear strain by this ratio, so as to compensate for the different forcing amplitude.
This implies that, for $A_w$ and $\lambda$ kept constant (same wave shape), increasing $\theta$ results in a decrease in the forcing amplitude, at least for angles in the range $\theta \in [50^\circ, 70^\circ]$.

\subsection{Streamwise velocity and Reynolds stresses}\label{sec:velRS}
\begin{figure*}
\includegraphics[scale=\scalefig]{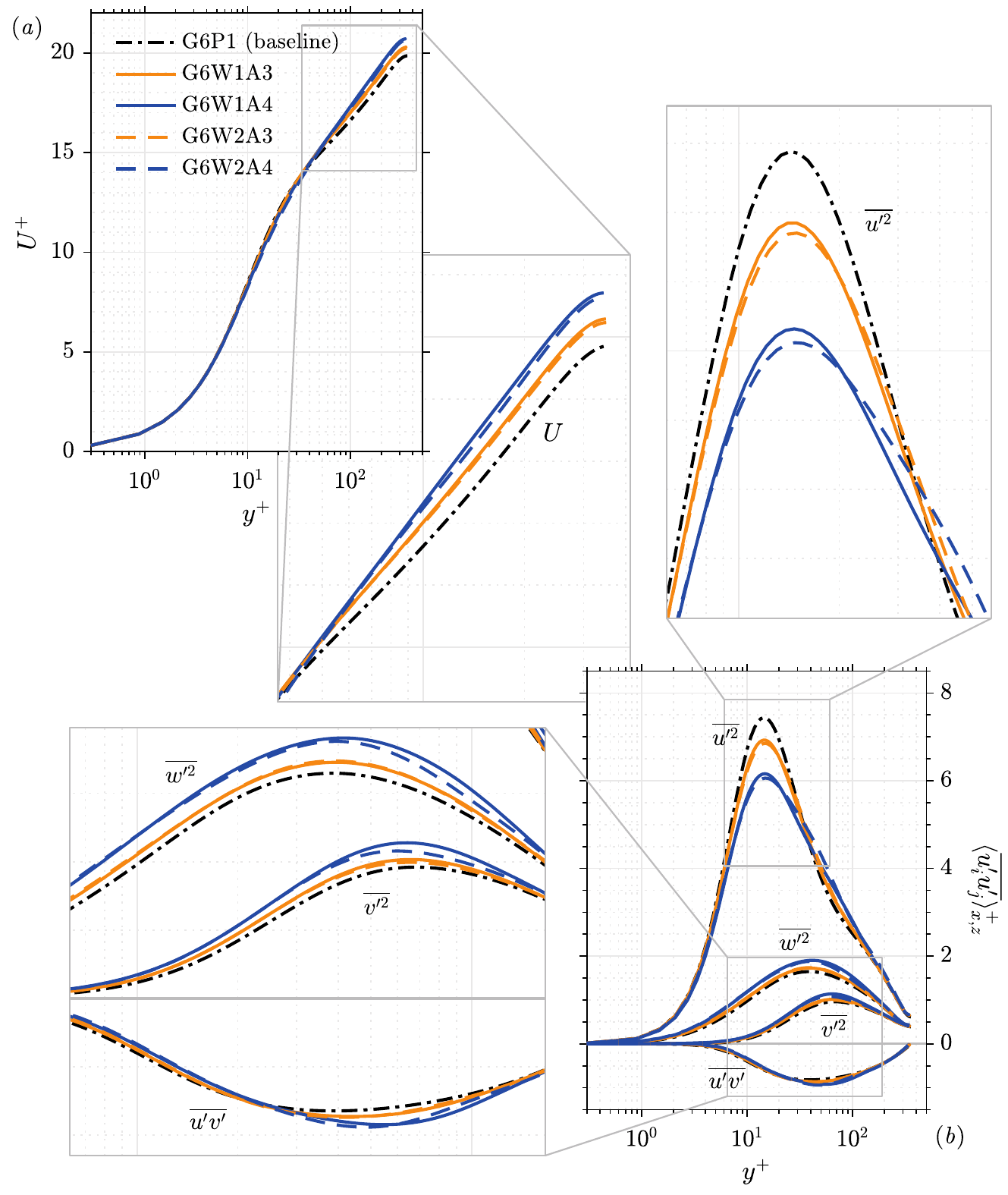}
\caption{Comparison of (\textit{a}) the mean streamwise velocity $U^+ = \langle \overline{u}\rangle_{x,z}/u_\tau$ and (\textit{b}) Reynolds stresses $\langle \overline{u'_iu'_j}\rangle_{x,z}/u_\tau^2$ for wavy walls with wave slopes of $2 A_w/\lambda \in \lbrace 0.05,0.07\rbrace$ (corresponding to labels *A3 and *A4) and wavelengths of $\lambda^+ \in \lbrace 612,918\rbrace$ (corresponding to labels *W2* and *W1*), and the baseline plane channel (G6P1) (scaled using the actual friction velocity).}
\label{fig:logRS}
\end{figure*}

As is observed with Stokes layers and, more generally, with most control strategies yielding turbulent-drag reduction, an upward shift in the log law appears relative to the uncontrolled flow when actual scaling is used (i.e. with the modified friction velocity).
This shift is shown in~\cref{fig:logRS}\,(\textit{a}) for some key simulations.
The corresponding flow configurations are characterised by two height-to-wavelength ratios and two wavelengths.
Although there is a noticeable difference between the two profiles for the two wavelengths, it is observed that similar wave slopes yield approximately similar shifts, the configuration $\lambda^+=918$ featuring a slightly greater shift that implies a higher level of skin-friction reduction (cf.~\cref{sec:FDR}).
As the wave slope is linked to the forcing amplitude, the higher this ratio is, the higher the friction-drag reduction is.
This close resemblance of the log shift at a given wave slope, but different wavelength, indicates a limited sensitivity of the friction-drag reduction for the range of wavelengths tested, especially at such a small forcing amplitude.

As far as the Reynolds stresses are concerned, there is a substantial decline in the peak of the streamwise normal stresses --~evidence of the weakening of the streaks.
The behaviour of the Reynolds stresses is also similar for a given height-to-wavelength ratio, apart from a detrimental increase in the streamwise normal stresses for the shorter wavelength ${\lambda^+ = 612}$, starting within the buffer layer around ${y^+ \approx 20}$ and persisting up to ${y^+\approx 80}$.
However, unlike wall-actuated Stokes layers \citep{Agostini2014,Touber2012,Viotti2009}, which entail a larger forcing amplitude, the decline in the streamwise stress only persists in the present configuration up to $y^+ \approx 35$, beyond which it exceeds the baseline value.
Similarly, the shear stress is depressed up to about the same wall-normal location, beyond which it also exceeds the baseline level.

\begin{figure*}\centering
\includegraphics[scale=\scalefig]{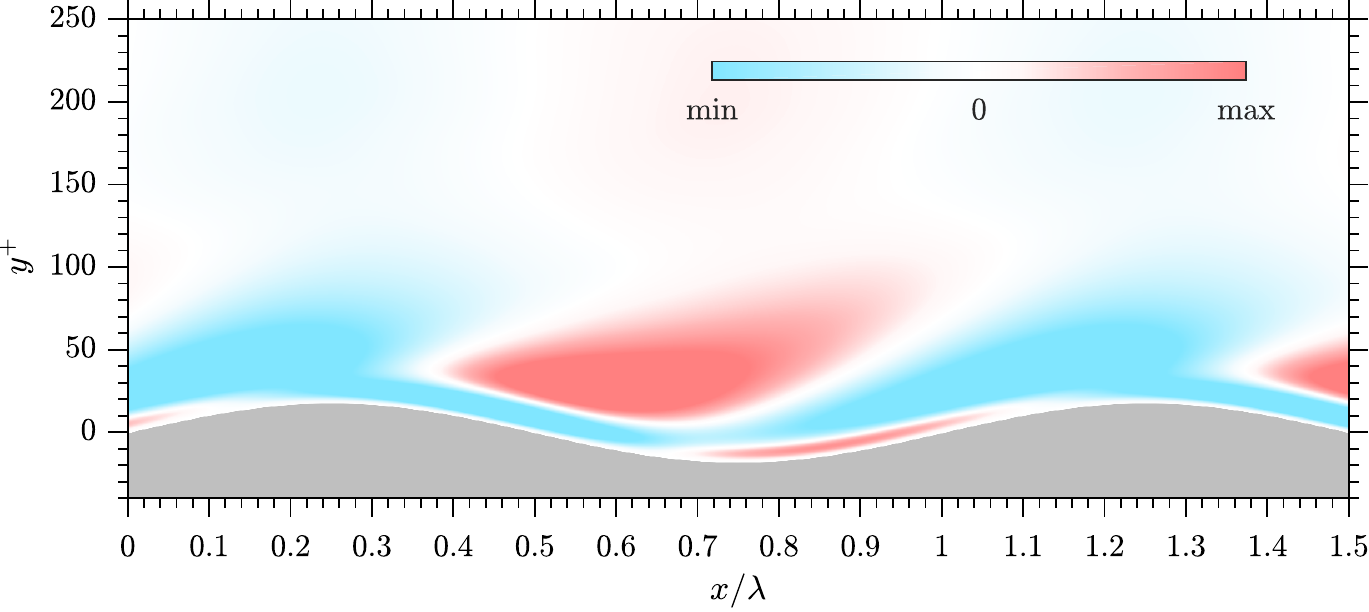}
\caption{(Colour online) Change in the streamwise Reynolds-stress component with respect to the baseline. Contours of $\widehat{u'^2}$ for G6W1A2 (${A_w^+=18}$, ${\theta=70^\circ}$, ${\lambda^+=918}$).}
\label{fig:uup}
\end{figure*}

The difference between the streamwise Reynolds-stress levels for the wavy walls and the baseline case is shown in~\cref{fig:uup} for G6W2A4. 
This brings to light two different regions showing distinct physical features.
One region is close to the wall, where a material weakening of the streaks takes place, and another is further away above the trough, featuring enhanced streamwise turbulence intensity.
The latter increase is stronger than the reduction above the crest, thus leading to a net increase in the mean streamwise turbulence intensity above ${y^+\approx 35}$, relative to the baseline (cf.~\cref{fig:logRS}\textit{b}).

\subsection{Detrimental effects of the wavy wall}\label{sec:ssldiff}
Despite the similarity between the shear-strain phase variations of wavy walls and Stokes layers, highlighted in~\cref{sec:shearcmp}, the effectiveness of the wavy wall is found to be lower.
In order to understand the lower performance, two main mechanisms are considered as possible causes of the degradation.

\begin{figure}\centering
\includegraphics[width=\linewidth]{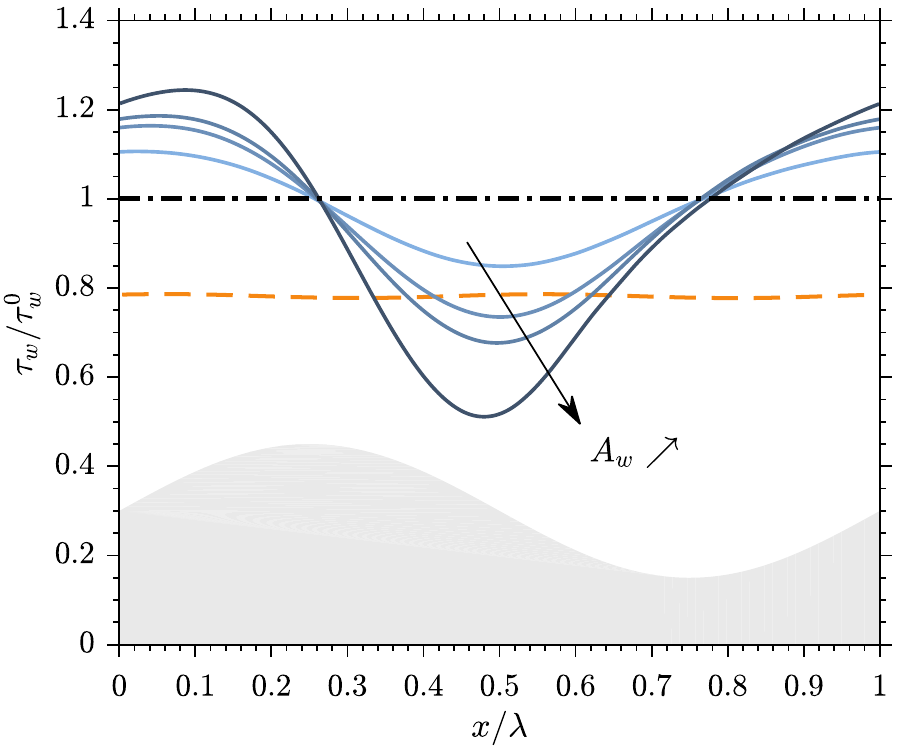}
\caption{Phase-wise variation of the turbulent skin friction. Thick dash-dotted line: plane channel, dashed line: SSL ($\lambda_x^+\approx 1250$, $A_\text{SSL}^+=2$), continuous lines: G6W1A*, i.e. for $\lambda^+ =918$, $\theta=70^\circ$, and $A_w^+\in\lbrace 11, 18, 22, 32 \rbrace$. The grey area shows the corresponding phase location on the wavy wall.}
\label{fig:tauw}
\end{figure}

First, beyond the observations made in~\cref{sec:velRS}, an important difference is that the friction is increased on the windward side of the wave, as shown in \cref{fig:tauw}.
The overall skin-friction reduction arises as a balance between the depressed friction on the leeward side of the wave and the enhanced friction on the windward side, whereas in the SSL case, the friction is decreased at all phases.
This variation is associated with the magnitude of the phase-varying streamwise velocity $\widetilde{u}$, which is greater than $\widetilde{w}$, whereas the former is almost negligible in the SSL (in optimum actuation conditions).
This phase-variation of the mean longitudinal velocity was already identified in \cite{Chernyshenko2013} as the main source of degradation of the performance of the wavy wall relative to that of the SSL.

\begin{figure*}
\begin{flushleft}
\vspace{10pt}\hspace{2.5cm}(\textit{a})\\ \vspace{-25pt}
\end{flushleft}
{\includegraphics[scale=\scalefig]{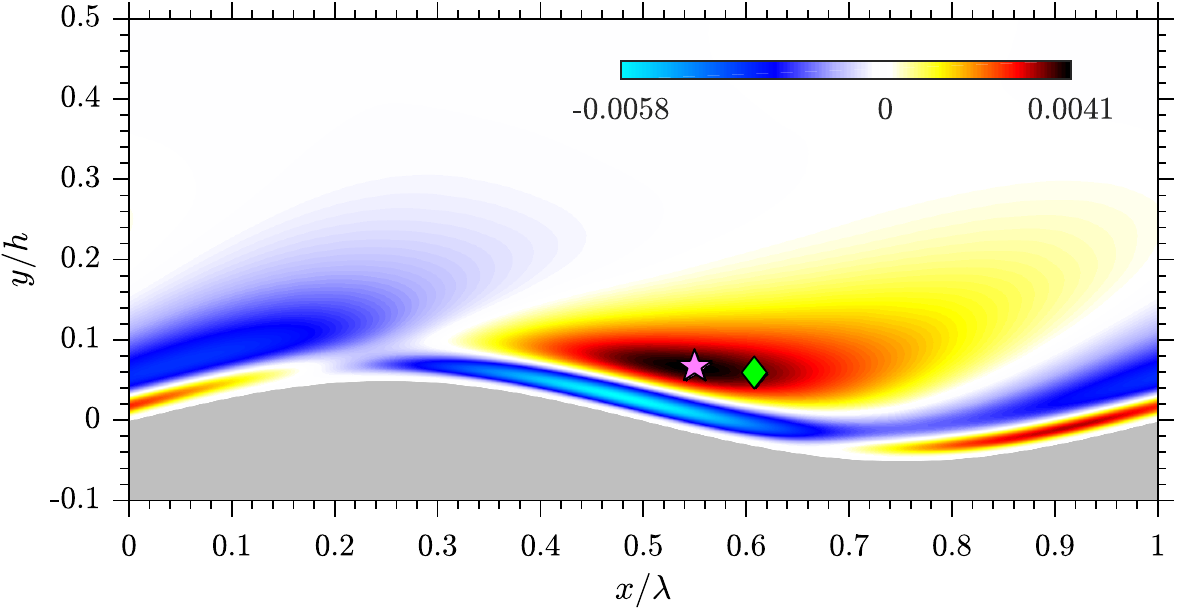}}\\
\begin{flushleft}
\vspace{10pt}\hspace{2.5cm}(\textit{b})\\ \vspace{-25pt}
\end{flushleft}
{\includegraphics[scale=\scalefig]{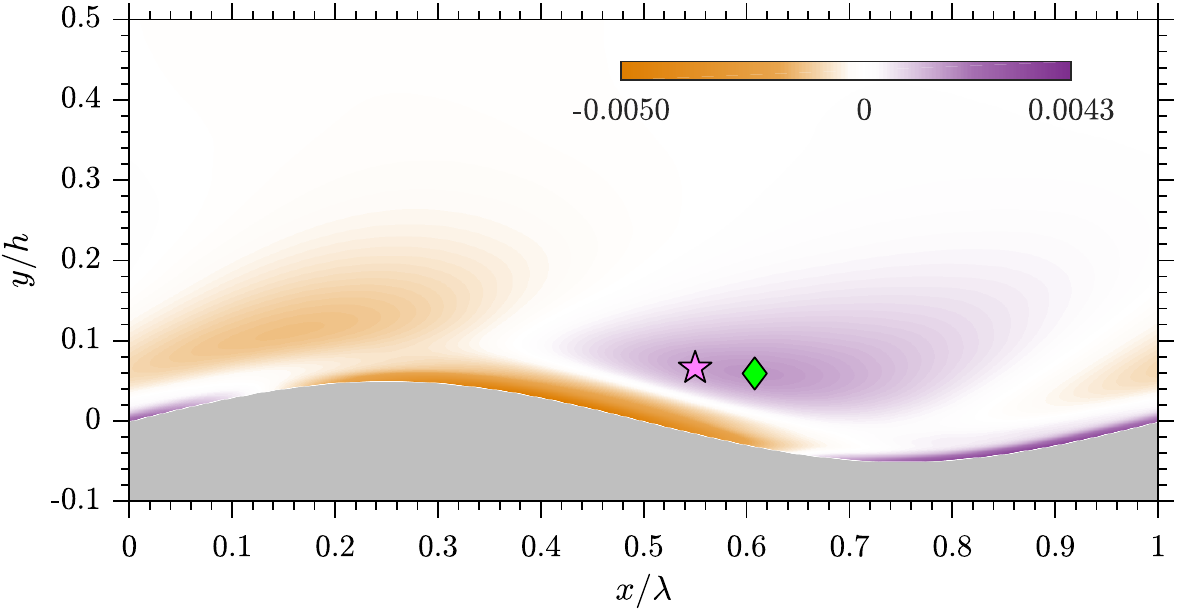}}%

\caption{Change in the production $\widehat{\Pi}_k$ and dissipation $-\widehat{\epsilon}_k$ from the baseline for G6W1A2 (${A_w^+=18}$, ${\theta=70^\circ}$, ${\lambda^+=918}$). (\textit{a})~production difference (\textit{b})~dissipation difference, $\bigstar$ peak in production difference, $\blacklozenge$ peak in dissipation difference.}
\label{fig:proddissip}
\end{figure*}

Second, an additional mechanism, specific to the wavy wall, was revealed by the numerical calculations.
As shown in~\cref{fig:proddissip}(\textit{a}), there exists a zone of intense production of turbulent kinetic energy ${\Pi_k = -\,\overline{u'_i u'_k} \,\partial \overline{u}_i/\partial x_k }$ above the leeward side of the wave, reflecting a deeper penetration of the disturbance arising from the wavy wall into the boundary layer.
As shown in \cref{fig:proddissip}(\textit{b}), the increase in production relative to the baseline is quickly followed, in phase, by an increase in energy dissipation ${\epsilon_k = -\nu\,\overline{\partial u'_i/\partial x_k\, \partial u'_i/\partial x_k}}$, at about the same wall-normal location.
This phenomenon strengthens as the wave amplitude increases, and is stronger for G6W2* than for G6W1* at similar wave slopes.

\subsection{Reynolds-number effect}

An interesting question is whether the flow properties remain similar when the shape of the wall is kept constant in viscous units as the Reynolds number is increased.
This has been investigated by reference to the flow configurations listed in~\cref{tab:Re}.

\begin{table}
  \centering
  \def~{\hphantom{0}}
	\begin{minipage}{0.5\textwidth}
	\begin{ruledtabular}
    \begin{tabular}{lcccc}
    Simulation label      & $\Rey_\tau$ & $A_w^+$ & $\theta$ & $\lambda^+$ \\
    \hline
    ~G1W1bis & 180   & 20    & 70$^\circ$    & 918 \\
    ~G2W1bis & 360   & 18    & 70$^\circ$    & 918 \\
    ~G5W1  & 1000  & 20    & 70$^\circ$    & 900 \\
    \end{tabular}%
    \end{ruledtabular}
    \end{minipage}
  \caption{Flow configurations for friction Reynolds numbers ranging from 180 to 1000.}
  \label{tab:Re}%
\end{table}%

\begin{figure*}\centering
\includegraphics[scale=\scalefig]{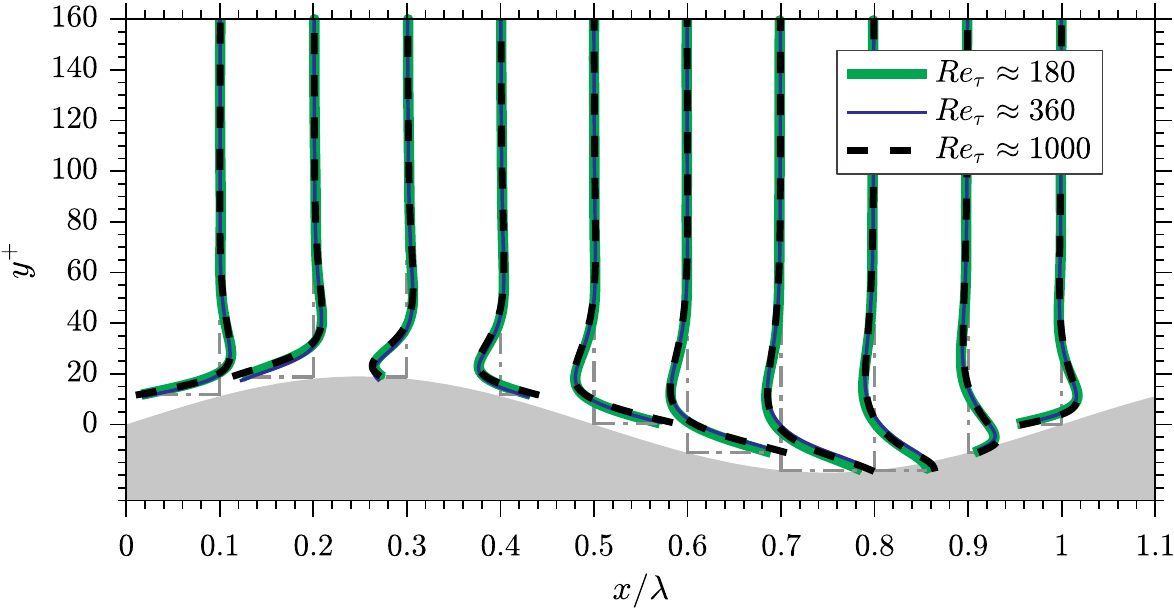}
\caption{Comparison of the shear strain $\partial \overline w^+ /\partial y^+$ for the same configuration at various Reynolds numbers. $A_w^+ \approx 20$, $\lambda^+\approx 900$, $\theta = 70^\circ$ (cf.~\cref{tab:Re}).}
\label{fig:reeffect}
\end{figure*}

Despite the wavy geometry in wall units not being exactly the same across the three flows in \cref{tab:Re}, and the mesh being somewhat coarser for the $\Rey_\tau = 1000$ case, the shear-strain profiles, scaled in wall units, are almost identical as demonstrated in \cref{fig:reeffect}.
At $\Rey_\tau = 180$, a wave height of $A_w^+ = 20$ represents a ratio $A_w/h$ of about 11\%, which is significant, while it decreases to 2\% for $\Rey_\tau = 1000$.
Consequently, the wave height $A_w$ is small enough relative to the channel height, so that, even at the lowest Reynolds number where the ratio $A_w/h$ is maximum, the distance between the walls remains sufficient to avoid an interference between the two solid boundaries, adding support to the findings reported in~\cref{sec:flat}.

Although net-drag-reduction levels of about 1--2\% were observed for the three Reynolds numbers tested, the value is not quantifiable with any degree of precision, because it is extremely sensitive to various numerical issues, as demonstrated in \cref{sec:gridcv} through a grid-convergence study at ${\Rey_\tau = 360}$.

\section{Conclusions}
In the present study, skewed wavy-wall channels have been investigated by means of direct numerical simulation as a potential passive open-loop drag-reduction device.
The spanwise-shear profiles generated by the wavy wall were shown to resemble closely those of the well-established method of drag reduction by in-plane wall motions, and to depend only weakly on the Reynolds number when expressed in wall units.
Various wavy-wall geometries for different combinations of flow angle $\theta$, wavelength $\lambda$, and height $A_w$ were explored with the aim of seeking a configuration that minimises the total drag relative to a plane wall.

Unlike for the Stokes layer, there is no actuation power required, but the skin-friction reduction is militated against by the pressure drag arising from the wavy geometry.
This drag increases quadratically with wave height, rapidly exceeding the friction-drag reduction beyond a modest wave height.
Consequently, the cumulative effect on the drag is small.
The corresponding accuracy requirements of the simulations, in terms of grid density and integration time, are, therefore, far beyond those generally-adopted in DNS of channel flows, making the cost of optimising the wavy-wall parameters prohibitive.
Even if relative changes in drag are quantified by reference to drag levels of a baseline plane channel, simulated with similar grid spacing, domain size, and flow-to-mesh angle, the net drag-reduction level is subject to a not insignificant error.

Despite the above qualifications, a net drag-reduction value of about 0.7\% (made up of a 2\%-friction reduction and a pressure-drag penalty of 1.3\%) was estimated for the configuration ${A_w^+ \approx 20}$, ${\theta = 70^\circ}$, ${\lambda^+ \approx 920}$, at $Re_\tau \approx 360$,  at the finest grid resolution, with indications that this performance would drop to 0.5\% for an asymptotically fine mesh.

The generation of a significant phase variation of the mean longitudinal velocity, proposed in earlier studies as a mechanism accounting for the degradation of the performance of the wavy wall relative to the steady Stokes layer, was augmented by the identification of a new mechanism consisting of the intense localised production of turbulence kinetic energy above the leeward side of the wave.


\begin{acknowledgements}
This study was funded by Innovate UK (Technology Strategy Board), as part of the ALFET project, project reference 113022.
The authors are grateful to the UK Turbulence Consortium (UKTC) for providing computational resources on the national supercomputing facility ARCHER under the EPSRC grant EP/L000261/1.
Access to Imperial College High Performance Computing Service, doi: 10.14469/hpc/2232 is also acknowledged.
\end{acknowledgements}

\bibliography{../../../cleanbib}

\begin{thebibliography}{39}%
\makeatletter
\providecommand \@ifxundefined [1]{%
 \@ifx{#1\undefined}
}%
\providecommand \@ifnum [1]{%
 \ifnum #1\expandafter \@firstoftwo
 \else \expandafter \@secondoftwo
 \fi
}%
\providecommand \@ifx [1]{%
 \ifx #1\expandafter \@firstoftwo
 \else \expandafter \@secondoftwo
 \fi
}%
\providecommand \natexlab [1]{#1}%
\providecommand \enquote  [1]{``#1''}%
\providecommand \bibnamefont  [1]{#1}%
\providecommand \bibfnamefont [1]{#1}%
\providecommand \citenamefont [1]{#1}%
\providecommand \href@noop [0]{\@secondoftwo}%
\providecommand \href [0]{\begingroup \@sanitize@url \@href}%
\providecommand \@href[1]{\@@startlink{#1}\@@href}%
\providecommand \@@href[1]{\endgroup#1\@@endlink}%
\providecommand \@sanitize@url [0]{\catcode `\\12\catcode `\$12\catcode
  `\&12\catcode `\#12\catcode `\^12\catcode `\_12\catcode `\%12\relax}%
\providecommand \@@startlink[1]{}%
\providecommand \@@endlink[0]{}%
\providecommand \url  [0]{\begingroup\@sanitize@url \@url }%
\providecommand \@url [1]{\endgroup\@href {#1}{\urlprefix }}%
\providecommand \urlprefix  [0]{URL }%
\providecommand \Eprint [0]{\href }%
\providecommand \doibase [0]{http://dx.doi.org/}%
\providecommand \selectlanguage [0]{\@gobble}%
\providecommand \bibinfo  [0]{\@secondoftwo}%
\providecommand \bibfield  [0]{\@secondoftwo}%
\providecommand \translation [1]{[#1]}%
\providecommand \BibitemOpen [0]{}%
\providecommand \bibitemStop [0]{}%
\providecommand \bibitemNoStop [0]{.\EOS\space}%
\providecommand \EOS [0]{\spacefactor3000\relax}%
\providecommand \BibitemShut  [1]{\csname bibitem#1\endcsname}%
\let\auto@bib@innerbib\@empty
\bibitem [{\citenamefont {Boomsma}\ and\ \citenamefont
  {Sotiropoulos}(2016)}]{Boomsma2016}%
  \BibitemOpen
  \bibfield  {author} {\bibinfo {author} {\bibfnamefont {A.}~\bibnamefont
  {Boomsma}}\ and\ \bibinfo {author} {\bibfnamefont {F.}~\bibnamefont
  {Sotiropoulos}},\ }\href {\doibase 10.1063/1.4942474} {\bibfield  {journal}
  {\bibinfo  {journal} {Phys. Fluids}\ }\textbf {\bibinfo {volume} {28}}
  (\bibinfo {year} {2016}),\ 10.1063/1.4942474}\BibitemShut {NoStop}%
\bibitem [{\citenamefont {Choi}\ \emph {et~al.}(1993)\citenamefont {Choi},
  \citenamefont {Moin},\ and\ \citenamefont {Kim}}]{Choi1993}%
  \BibitemOpen
  \bibfield  {author} {\bibinfo {author} {\bibfnamefont {H.}~\bibnamefont
  {Choi}}, \bibinfo {author} {\bibfnamefont {P.}~\bibnamefont {Moin}}, \ and\
  \bibinfo {author} {\bibfnamefont {J.}~\bibnamefont {Kim}},\ }\href@noop {}
  {\bibfield  {journal} {\bibinfo  {journal} {J. Fluid Mech.}\ }\textbf
  {\bibinfo {volume} {255}},\ \bibinfo {pages} {503} (\bibinfo {year}
  {1993})}\BibitemShut {NoStop}%
\bibitem [{\citenamefont {Bechert}\ and\ \citenamefont
  {Bartenwerfer}(1989)}]{Bechert1989}%
  \BibitemOpen
  \bibfield  {author} {\bibinfo {author} {\bibfnamefont {D.~W.}\ \bibnamefont
  {Bechert}}\ and\ \bibinfo {author} {\bibfnamefont {M.}~\bibnamefont
  {Bartenwerfer}},\ }\href@noop {} {\bibfield  {journal} {\bibinfo  {journal}
  {J. Fluid Mech.}\ }\textbf {\bibinfo {volume} {296}},\ \bibinfo {pages} {105}
  (\bibinfo {year} {1989})}\BibitemShut {NoStop}%
\bibitem [{\citenamefont {Bechert}\ \emph {et~al.}(1997)\citenamefont
  {Bechert}, \citenamefont {Bruse}, \citenamefont {Hage}, \citenamefont
  {van~der Hoeven},\ and\ \citenamefont {Hoppe}}]{Bechert1997}%
  \BibitemOpen
  \bibfield  {author} {\bibinfo {author} {\bibfnamefont {D.~W.}\ \bibnamefont
  {Bechert}}, \bibinfo {author} {\bibfnamefont {M.}~\bibnamefont {Bruse}},
  \bibinfo {author} {\bibfnamefont {W.}~\bibnamefont {Hage}}, \bibinfo {author}
  {\bibfnamefont {J.~G.~T.}\ \bibnamefont {van~der Hoeven}}, \ and\ \bibinfo
  {author} {\bibfnamefont {G.}~\bibnamefont {Hoppe}},\ }\href {\doibase
  10.1017/S0022112096004673} {\bibfield  {journal} {\bibinfo  {journal} {J.
  Fluid Mech.}\ }\textbf {\bibinfo {volume} {338}},\ \bibinfo {pages} {59}
  (\bibinfo {year} {1997})}\BibitemShut {NoStop}%
\bibitem [{\citenamefont {Garc\'ia-Mayoral}\ and\ \citenamefont
  {Jim\'enez}(2011)}]{Garcia-Mayoral2011}%
  \BibitemOpen
  \bibfield  {author} {\bibinfo {author} {\bibfnamefont {R.}~\bibnamefont
  {Garc\'ia-Mayoral}}\ and\ \bibinfo {author} {\bibfnamefont {J.}~\bibnamefont
  {Jim\'enez}},\ }\href {\doibase 10.1098/rsta.2010.0359} {\bibfield  {journal}
  {\bibinfo  {journal} {Philosophical Transactions of the Royal Society A}\
  }\textbf {\bibinfo {volume} {369}},\ \bibinfo {pages} {1412} (\bibinfo {year}
  {2011})}\BibitemShut {NoStop}%
\bibitem [{\citenamefont {Peet}\ \emph {et~al.}(2009)\citenamefont {Peet},
  \citenamefont {Sagaut},\ and\ \citenamefont {Charron}}]{Peet2009b}%
  \BibitemOpen
  \bibfield  {author} {\bibinfo {author} {\bibfnamefont {Y.}~\bibnamefont
  {Peet}}, \bibinfo {author} {\bibfnamefont {P.}~\bibnamefont {Sagaut}}, \ and\
  \bibinfo {author} {\bibfnamefont {Y.}~\bibnamefont {Charron}},\ }\href@noop
  {} {\bibfield  {journal} {\bibinfo  {journal} {Int. J. Hydrogen Energy}\
  }\textbf {\bibinfo {volume} {34}},\ \bibinfo {pages} {8964} (\bibinfo {year}
  {2009})}\BibitemShut {NoStop}%
\bibitem [{\citenamefont {Kramer}\ \emph {et~al.}(2010)\citenamefont {Kramer},
  \citenamefont {Gru\"uneberger}, \citenamefont {Thiele},\ and\ \citenamefont
  {Wassen}}]{Kramer2010}%
  \BibitemOpen
  \bibfield  {author} {\bibinfo {author} {\bibfnamefont {F.}~\bibnamefont
  {Kramer}}, \bibinfo {author} {\bibfnamefont {R.}~\bibnamefont
  {Gru\"uneberger}}, \bibinfo {author} {\bibfnamefont {F.}~\bibnamefont
  {Thiele}}, \ and\ \bibinfo {author} {\bibfnamefont {E.}~\bibnamefont
  {Wassen}},\ }in\ \href@noop {} {\emph {\bibinfo {booktitle} {5th Flow control
  Conference}}},\ \bibinfo {series and number} {AIAA 2010-4583}\ (\bibinfo
  {year} {2010})\BibitemShut {NoStop}%
\bibitem [{\citenamefont {Bannier}(2016)}]{BannierPhD}%
  \BibitemOpen
  \bibfield  {author} {\bibinfo {author} {\bibfnamefont {A.}~\bibnamefont
  {Bannier}},\ }\emph {\bibinfo {title} {Contr\^ole de la tra\^in\'ee de
  frottement d'une couche limite turbulente au moyen de rev\^etements
  rainur\'es de type riblets}},\ \href@noop {} {\bibinfo {type} {M\'ecanique
  des fluides [physics.class-ph]}},\ \bibinfo  {school} {Universit\'e Pierre et
  Marie Curie, Paris VI} (\bibinfo {year} {2016})\BibitemShut {NoStop}%
\bibitem [{\citenamefont {Jung}\ \emph {et~al.}(1992)\citenamefont {Jung},
  \citenamefont {Mangiavacchi},\ and\ \citenamefont {Akhavan}}]{Jung1992}%
  \BibitemOpen
  \bibfield  {author} {\bibinfo {author} {\bibfnamefont {W.~J.}\ \bibnamefont
  {Jung}}, \bibinfo {author} {\bibfnamefont {N.}~\bibnamefont {Mangiavacchi}},
  \ and\ \bibinfo {author} {\bibfnamefont {R.}~\bibnamefont {Akhavan}},\
  }\href@noop {} {\bibfield  {journal} {\bibinfo  {journal} {Phys. Fluids}\
  }\textbf {\bibinfo {volume} {4}},\ \bibinfo {pages} {1605} (\bibinfo {year}
  {1992})}\BibitemShut {NoStop}%
\bibitem [{\citenamefont {Quadrio}\ \emph {et~al.}(2009)\citenamefont
  {Quadrio}, \citenamefont {Ricco},\ and\ \citenamefont
  {Viotti}}]{Quadrio2009}%
  \BibitemOpen
  \bibfield  {author} {\bibinfo {author} {\bibfnamefont {M.}~\bibnamefont
  {Quadrio}}, \bibinfo {author} {\bibfnamefont {P.}~\bibnamefont {Ricco}}, \
  and\ \bibinfo {author} {\bibfnamefont {C.}~\bibnamefont {Viotti}},\
  }\href@noop {} {\bibfield  {journal} {\bibinfo  {journal} {J. Fluid Mech.}\
  }\textbf {\bibinfo {volume} {627}},\ \bibinfo {pages} {161} (\bibinfo {year}
  {2009})}\BibitemShut {NoStop}%
\bibitem [{\citenamefont {Quadrio}\ and\ \citenamefont
  {Ricco}(2010)}]{Quadrio2010}%
  \BibitemOpen
  \bibfield  {author} {\bibinfo {author} {\bibfnamefont {M.}~\bibnamefont
  {Quadrio}}\ and\ \bibinfo {author} {\bibfnamefont {P.}~\bibnamefont
  {Ricco}},\ }\href {\doibase 10.1017/S0022112010004398} {\bibfield  {journal}
  {\bibinfo  {journal} {J. Fluid Mech.}\ }\textbf {\bibinfo {volume} {667}},\
  \bibinfo {pages} {135} (\bibinfo {year} {2010})}\BibitemShut {NoStop}%
\bibitem [{\citenamefont {Touber}\ and\ \citenamefont
  {Leschziner}(2012)}]{Touber2012}%
  \BibitemOpen
  \bibfield  {author} {\bibinfo {author} {\bibfnamefont {E.}~\bibnamefont
  {Touber}}\ and\ \bibinfo {author} {\bibfnamefont {M.~A.}\ \bibnamefont
  {Leschziner}},\ }\href {\doibase 10.1017/jfm.2011.507} {\bibfield  {journal}
  {\bibinfo  {journal} {J. Fluid Mech.}\ }\textbf {\bibinfo {volume} {693}},\
  \bibinfo {pages} {150} (\bibinfo {year} {2012})}\BibitemShut {NoStop}%
\bibitem [{\citenamefont {Hurst}\ \emph {et~al.}(2014)\citenamefont {Hurst},
  \citenamefont {Yang},\ and\ \citenamefont {Chung}}]{Hurst2014}%
  \BibitemOpen
  \bibfield  {author} {\bibinfo {author} {\bibfnamefont {E.}~\bibnamefont
  {Hurst}}, \bibinfo {author} {\bibfnamefont {Q.}~\bibnamefont {Yang}}, \ and\
  \bibinfo {author} {\bibfnamefont {Y.~M.}\ \bibnamefont {Chung}},\ }\href
  {\doibase 10.1017/jfm.2014.524} {\bibfield  {journal} {\bibinfo  {journal}
  {J. Fluid Mech.}\ }\textbf {\bibinfo {volume} {759}},\ \bibinfo {pages} {28}
  (\bibinfo {year} {2014})}\BibitemShut {NoStop}%
\bibitem [{\citenamefont {Gatti}\ and\ \citenamefont
  {Quadrio}(2016)}]{Gatti2016}%
  \BibitemOpen
  \bibfield  {author} {\bibinfo {author} {\bibfnamefont {D.}~\bibnamefont
  {Gatti}}\ and\ \bibinfo {author} {\bibfnamefont {M.}~\bibnamefont
  {Quadrio}},\ }\href {\doibase 10.1017/jfm.2016.485} {\bibfield  {journal}
  {\bibinfo  {journal} {J. Fluid Mech.}\ }\textbf {\bibinfo {volume} {802}},\
  \bibinfo {pages} {553} (\bibinfo {year} {2016})}\BibitemShut {NoStop}%
\bibitem [{\citenamefont {Auteri}\ \emph {et~al.}(2010)\citenamefont {Auteri},
  \citenamefont {Baron}, \citenamefont {Belan}, \citenamefont {Campanardi},\
  and\ \citenamefont {Quadrio}}]{Auteri2010}%
  \BibitemOpen
  \bibfield  {author} {\bibinfo {author} {\bibfnamefont {F.}~\bibnamefont
  {Auteri}}, \bibinfo {author} {\bibfnamefont {A.}~\bibnamefont {Baron}},
  \bibinfo {author} {\bibfnamefont {M.}~\bibnamefont {Belan}}, \bibinfo
  {author} {\bibfnamefont {G.}~\bibnamefont {Campanardi}}, \ and\ \bibinfo
  {author} {\bibfnamefont {M.}~\bibnamefont {Quadrio}},\ }\href@noop {}
  {\bibfield  {journal} {\bibinfo  {journal} {Phys. Fluids}\ }\textbf {\bibinfo
  {volume} {22}} (\bibinfo {year} {2010})}\BibitemShut {NoStop}%
\bibitem [{\citenamefont {Bird}\ \emph {et~al.}(2015)\citenamefont {Bird},
  \citenamefont {Santer},\ and\ \citenamefont {Morrison}}]{Bird2015}%
  \BibitemOpen
  \bibfield  {author} {\bibinfo {author} {\bibfnamefont {J.}~\bibnamefont
  {Bird}}, \bibinfo {author} {\bibfnamefont {M.}~\bibnamefont {Santer}}, \ and\
  \bibinfo {author} {\bibfnamefont {J.~F.}\ \bibnamefont {Morrison}},\ }in\
  \href@noop {} {\emph {\bibinfo {booktitle} {European Drag Reduction and Flow
  Control Meeting, Cambridge, UK}}}\ (\bibinfo {organization} {ERCOFTAC},\
  \bibinfo {year} {2015})\BibitemShut {NoStop}%
\bibitem [{\citenamefont {Viotti}\ \emph {et~al.}(2009)\citenamefont {Viotti},
  \citenamefont {Quadrio},\ and\ \citenamefont {Luchini}}]{Viotti2009}%
  \BibitemOpen
  \bibfield  {author} {\bibinfo {author} {\bibfnamefont {C.}~\bibnamefont
  {Viotti}}, \bibinfo {author} {\bibfnamefont {M.}~\bibnamefont {Quadrio}}, \
  and\ \bibinfo {author} {\bibfnamefont {P.}~\bibnamefont {Luchini}},\ }\href
  {\doibase 10.1063/1.3266945} {\bibfield  {journal} {\bibinfo  {journal}
  {Phys. Fluids}\ }\textbf {\bibinfo {volume} {21}} (\bibinfo {year} {2009}),\
  10.1063/1.3266945}\BibitemShut {NoStop}%
\bibitem [{\citenamefont {Chernyshenko}(2013)}]{Chernyshenko2013}%
  \BibitemOpen
  \bibfield  {author} {\bibinfo {author} {\bibfnamefont {S.}~\bibnamefont
  {Chernyshenko}},\ }\href@noop {} {\bibfield  {journal} {\bibinfo  {journal}
  {ArXiv e-prints}\ }\textbf {\bibinfo {volume} {[physics.flu-dyn]}} (\bibinfo
  {year} {2013})},\ \Eprint {http://arxiv.org/abs/1204.4638} {arXiv:1204.4638
  [physics.flu-dyn]} \BibitemShut {NoStop}%
\bibitem [{\citenamefont {Fishpool}\ and\ \citenamefont
  {Leschziner}(2009)}]{Fishpool2009}%
  \BibitemOpen
  \bibfield  {author} {\bibinfo {author} {\bibfnamefont {G.~M.}\ \bibnamefont
  {Fishpool}}\ and\ \bibinfo {author} {\bibfnamefont {M.~A.}\ \bibnamefont
  {Leschziner}},\ }\href@noop {} {\bibfield  {journal} {\bibinfo  {journal}
  {Computer \& Fluids}\ }\textbf {\bibinfo {volume} {38}},\ \bibinfo {pages}
  {1289} (\bibinfo {year} {2009})}\BibitemShut {NoStop}%
\bibitem [{\citenamefont {Hirsch}(2007)}]{Hirsch2007}%
  \BibitemOpen
  \bibfield  {author} {\bibinfo {author} {\bibfnamefont {C.}~\bibnamefont
  {Hirsch}},\ }\href@noop {} {\emph {\bibinfo {title} {Numerical Computation of
  Internal and External Flows: Fundamentals of Computational Fluid
  Dynamics}}},\ \bibinfo {edition} {2nd}\ ed.\ (\bibinfo  {publisher}
  {Elsevier},\ \bibinfo {year} {2007})\BibitemShut {NoStop}%
\bibitem [{\citenamefont {Lien}\ and\ \citenamefont
  {Leschziner}(1993)}]{Lien1993}%
  \BibitemOpen
  \bibfield  {author} {\bibinfo {author} {\bibfnamefont {F.~S.}\ \bibnamefont
  {Lien}}\ and\ \bibinfo {author} {\bibfnamefont {M.~A.}\ \bibnamefont
  {Leschziner}},\ }\href@noop {} {\bibfield  {journal} {\bibinfo  {journal}
  {Computational Methods in Applied Mechanics and Engineering}\ }\textbf
  {\bibinfo {volume} {118}},\ \bibinfo {pages} {351} (\bibinfo {year}
  {1993})}\BibitemShut {NoStop}%
\bibitem [{\citenamefont {Rhie}\ and\ \citenamefont {Chow}(1983)}]{Rhie1983}%
  \BibitemOpen
  \bibfield  {author} {\bibinfo {author} {\bibfnamefont {C.~M.}\ \bibnamefont
  {Rhie}}\ and\ \bibinfo {author} {\bibfnamefont {W.~L.}\ \bibnamefont
  {Chow}},\ }\href {\doibase 10.2514/3.8284} {\bibfield  {journal} {\bibinfo
  {journal} {AIAA Journal}\ }\textbf {\bibinfo {volume} {21}},\ \bibinfo
  {pages} {1525} (\bibinfo {year} {1983})}\BibitemShut {NoStop}%
\bibitem [{\citenamefont {Roache}(1997)}]{Roache1997}%
  \BibitemOpen
  \bibfield  {author} {\bibinfo {author} {\bibfnamefont {P.~J.}\ \bibnamefont
  {Roache}},\ }\href {\doibase 10.1146/annurev.fluid.29.1.123} {\bibfield
  {journal} {\bibinfo  {journal} {Annu. Rev. Fluid Mech.}\ }\textbf {\bibinfo
  {volume} {29}},\ \bibinfo {pages} {123} (\bibinfo {year} {1997})}\BibitemShut
  {NoStop}%
\bibitem [{\citenamefont {Roache}(1998)}]{Roache1998}%
  \BibitemOpen
  \bibfield  {author} {\bibinfo {author} {\bibfnamefont {P.~J.}\ \bibnamefont
  {Roache}},\ }\href@noop {} {\emph {\bibinfo {title} {Verification and
  validation in computational science and engineering}}}\ (\bibinfo
  {publisher} {Hermosa Publishers},\ \bibinfo {year} {1998})\BibitemShut
  {NoStop}%
\bibitem [{\citenamefont {Roache}(2002)}]{Roache2002}%
  \BibitemOpen
  \bibfield  {author} {\bibinfo {author} {\bibfnamefont {P.~J.}\ \bibnamefont
  {Roache}},\ }\href {\doibase 10.1115/1.1436090} {\bibfield  {journal}
  {\bibinfo  {journal} {ASME journal of Fluids Engineering}\ }\textbf {\bibinfo
  {volume} {114}},\ \bibinfo {pages} {4} (\bibinfo {year} {2002})}\BibitemShut
  {NoStop}%
\bibitem [{\citenamefont {Roy}\ \emph {et~al.}(2004)\citenamefont {Roy},
  \citenamefont {Nelson}, \citenamefont {Smith},\ and\ \citenamefont
  {Ober}}]{Roy2004}%
  \BibitemOpen
  \bibfield  {author} {\bibinfo {author} {\bibfnamefont {C.~J.}\ \bibnamefont
  {Roy}}, \bibinfo {author} {\bibfnamefont {C.~C.}\ \bibnamefont {Nelson}},
  \bibinfo {author} {\bibfnamefont {T.~M.}\ \bibnamefont {Smith}}, \ and\
  \bibinfo {author} {\bibfnamefont {C.~C.}\ \bibnamefont {Ober}},\ }\href
  {\doibase 10.1002/fld.660} {\bibfield  {journal} {\bibinfo  {journal}
  {International Journal for Numerical Methods in Fluids}\ }\textbf {\bibinfo
  {volume} {44}},\ \bibinfo {pages} {599} (\bibinfo {year} {2004})}\BibitemShut
  {NoStop}%
\bibitem [{\citenamefont {Salari}\ and\ \citenamefont
  {Knupp}(2000)}]{Salari2000}%
  \BibitemOpen
  \bibfield  {author} {\bibinfo {author} {\bibfnamefont {K.}~\bibnamefont
  {Salari}}\ and\ \bibinfo {author} {\bibfnamefont {P.}~\bibnamefont {Knupp}},\
  }\href@noop {} {\emph {\bibinfo {title} {Code verification by the method of
  manufactured solutions}}}\ (\bibinfo  {publisher} {Sandia National
  Laboratories},\ \bibinfo {year} {2000})\BibitemShut {NoStop}%
\bibitem [{\citenamefont {Maa{\ss}}\ and\ \citenamefont
  {Schumann}(1996)}]{Maass1996}%
  \BibitemOpen
  \bibfield  {author} {\bibinfo {author} {\bibfnamefont {C.}~\bibnamefont
  {Maa{\ss}}}\ and\ \bibinfo {author} {\bibfnamefont {U.}~\bibnamefont
  {Schumann}},\ }in\ \href {\doibase 10.1007/978-3-322-89849-4_17} {\emph
  {\bibinfo {booktitle} {Flow Simulation with High-Performance Computers
  II}}},\ \bibinfo {series} {Notes on Numerical Fluid Mechanics (NNFM)},
  Vol.~\bibinfo {volume} {48},\ \bibinfo {editor} {edited by\ \bibinfo {editor}
  {\bibfnamefont {E.}~\bibnamefont {Hirschel}}}\ (\bibinfo  {publisher}
  {Vieweg+Teubner Verlag},\ \bibinfo {year} {1996})\ pp.\ \bibinfo {pages}
  {227--241}\BibitemShut {NoStop}%
\bibitem [{\citenamefont {Wang}\ \emph {et~al.}(2006)\citenamefont {Wang},
  \citenamefont {Yeo},\ and\ \citenamefont {Khoo}}]{Wang2006}%
  \BibitemOpen
  \bibfield  {author} {\bibinfo {author} {\bibfnamefont {Z.}~\bibnamefont
  {Wang}}, \bibinfo {author} {\bibfnamefont {K.~S.}\ \bibnamefont {Yeo}}, \
  and\ \bibinfo {author} {\bibfnamefont {B.~C.}\ \bibnamefont {Khoo}},\ }\href
  {\doibase 10.1080/14685240600595735} {\bibfield  {journal} {\bibinfo
  {journal} {Journal of Turbulence}\ }\textbf {\bibinfo {volume} {7}} (\bibinfo
  {year} {2006}),\ 10.1080/14685240600595735}\BibitemShut {NoStop}%
\bibitem [{\citenamefont {Cherukat}\ \emph {et~al.}(1998)\citenamefont
  {Cherukat}, \citenamefont {Na}, \citenamefont {Hanratty},\ and\ \citenamefont
  {McLaughlin}}]{Cherukat1998}%
  \BibitemOpen
  \bibfield  {author} {\bibinfo {author} {\bibfnamefont {P.}~\bibnamefont
  {Cherukat}}, \bibinfo {author} {\bibfnamefont {Y.}~\bibnamefont {Na}},
  \bibinfo {author} {\bibfnamefont {T.~J.}\ \bibnamefont {Hanratty}}, \ and\
  \bibinfo {author} {\bibfnamefont {J.~B.}\ \bibnamefont {McLaughlin}},\
  }\href@noop {} {\bibfield  {journal} {\bibinfo  {journal} {Theoretical
  Computational Fluid Dynamics}\ }\textbf {\bibinfo {volume} {11}},\ \bibinfo
  {pages} {109} (\bibinfo {year} {1998})}\BibitemShut {NoStop}%
\bibitem [{\citenamefont {Henn}\ and\ \citenamefont {Sykes}(1999)}]{Henn1999}%
  \BibitemOpen
  \bibfield  {author} {\bibinfo {author} {\bibfnamefont {D.~S.}\ \bibnamefont
  {Henn}}\ and\ \bibinfo {author} {\bibfnamefont {R.~I.}\ \bibnamefont
  {Sykes}},\ }\href {\doibase 10.1017/S0022112098003723} {\bibfield  {journal}
  {\bibinfo  {journal} {J. Fluid Mech.}\ }\textbf {\bibinfo {volume} {383}},\
  \bibinfo {pages} {75} (\bibinfo {year} {1999})}\BibitemShut {NoStop}%
\bibitem [{\citenamefont {Yoon}\ \emph {et~al.}(2009)\citenamefont {Yoon},
  \citenamefont {El-Samni}, \citenamefont {Huynh}, \citenamefont {Chun},
  \citenamefont {Kim}, \citenamefont {Pham},\ and\ \citenamefont
  {Park}}]{Yoon2009}%
  \BibitemOpen
  \bibfield  {author} {\bibinfo {author} {\bibfnamefont {H.}~\bibnamefont
  {Yoon}}, \bibinfo {author} {\bibfnamefont {O.~A.}\ \bibnamefont {El-Samni}},
  \bibinfo {author} {\bibfnamefont {A.}~\bibnamefont {Huynh}}, \bibinfo
  {author} {\bibfnamefont {H.~H.}\ \bibnamefont {Chun}}, \bibinfo {author}
  {\bibfnamefont {H.~J.}\ \bibnamefont {Kim}}, \bibinfo {author} {\bibfnamefont
  {A.~H.}\ \bibnamefont {Pham}}, \ and\ \bibinfo {author} {\bibfnamefont
  {I.~R.}\ \bibnamefont {Park}},\ }\href {\doibase
  10.1016/j.oceaneng.2009.03.012} {\bibfield  {journal} {\bibinfo  {journal}
  {Ocean Engineering}\ }\textbf {\bibinfo {volume} {36}},\ \bibinfo {pages}
  {697} (\bibinfo {year} {2009})}\BibitemShut {NoStop}%
\bibitem [{\citenamefont {Nakanishi}\ \emph {et~al.}(2012)\citenamefont
  {Nakanishi}, \citenamefont {Mamori},\ and\ \citenamefont
  {Fukagata}}]{Nakanishi2012}%
  \BibitemOpen
  \bibfield  {author} {\bibinfo {author} {\bibfnamefont {R.}~\bibnamefont
  {Nakanishi}}, \bibinfo {author} {\bibfnamefont {H.}~\bibnamefont {Mamori}}, \
  and\ \bibinfo {author} {\bibfnamefont {K.}~\bibnamefont {Fukagata}},\ }\href
  {\doibase 10.1016/j.ijheatfluidflow.2012.01.007} {\bibfield  {journal}
  {\bibinfo  {journal} {International Journal of Heat and Fluid Flow}\ }\textbf
  {\bibinfo {volume} {35}},\ \bibinfo {pages} {152} (\bibinfo {year}
  {2012})}\BibitemShut {NoStop}%
\bibitem [{\citenamefont {Mamori}(2012)}]{Mamori2012}%
  \BibitemOpen
  \bibfield  {author} {\bibinfo {author} {\bibfnamefont {H.}~\bibnamefont
  {Mamori}},\ }\emph {\bibinfo {title} {Numerical analysis of drag reduction in
  channel flow by traveling wave-like blowing and suction}},\ \href@noop {}
  {Ph.D. thesis},\ \bibinfo  {school} {Keio University} (\bibinfo {year}
  {2012})\BibitemShut {NoStop}%
\bibitem [{\citenamefont {Luchini}(2016)}]{Luchini2016}%
  \BibitemOpen
  \bibfield  {author} {\bibinfo {author} {\bibfnamefont {P.}~\bibnamefont
  {Luchini}},\ }\href {\doibase 10.1016/j.euromechflu.2015.08.007} {\bibfield
  {journal} {\bibinfo  {journal} {European Journal of Mechanics B/Fluids}\
  }\textbf {\bibinfo {volume} {55}},\ \bibinfo {pages} {340} (\bibinfo {year}
  {2016})}\BibitemShut {NoStop}%
\bibitem [{\citenamefont {Ricco}\ and\ \citenamefont {Wu}(2004)}]{Ricco2004}%
  \BibitemOpen
  \bibfield  {author} {\bibinfo {author} {\bibfnamefont {P.}~\bibnamefont
  {Ricco}}\ and\ \bibinfo {author} {\bibfnamefont {S.}~\bibnamefont {Wu}},\
  }\href {\doibase 10.1016/j.expthermflusci.2004.01.010} {\bibfield  {journal}
  {\bibinfo  {journal} {Experimental Thermal and Fluid and Science}\ }\textbf
  {\bibinfo {volume} {29}},\ \bibinfo {pages} {41} (\bibinfo {year}
  {2004})}\BibitemShut {NoStop}%
\bibitem [{\citenamefont {Gatti}\ and\ \citenamefont
  {Quadrio}(2013)}]{Gatti2013}%
  \BibitemOpen
  \bibfield  {author} {\bibinfo {author} {\bibfnamefont {D.}~\bibnamefont
  {Gatti}}\ and\ \bibinfo {author} {\bibfnamefont {M.}~\bibnamefont
  {Quadrio}},\ }\href {\doibase 10.1063/1.4849537} {\bibfield  {journal}
  {\bibinfo  {journal} {Phys. Fluids}\ }\textbf {\bibinfo {volume} {25}}
  (\bibinfo {year} {2013}),\ 10.1063/1.4849537}\BibitemShut {NoStop}%
\bibitem [{\citenamefont {Russo}\ and\ \citenamefont
  {Luchini}(2017)}]{Luchini2017}%
  \BibitemOpen
  \bibfield  {author} {\bibinfo {author} {\bibfnamefont {S.}~\bibnamefont
  {Russo}}\ and\ \bibinfo {author} {\bibfnamefont {P.}~\bibnamefont
  {Luchini}},\ }\href@noop {} {\enquote {\bibinfo {title} {A fast algorithm for
  the estimation of statistical error in {DNS} (or experimental) time
  averages},}\ } (\bibinfo {year} {2017})\BibitemShut {NoStop}%
\bibitem [{\citenamefont {Agostini}\ \emph {et~al.}(2014)\citenamefont
  {Agostini}, \citenamefont {Touber},\ and\ \citenamefont
  {Leschziner}}]{Agostini2014}%
  \BibitemOpen
  \bibfield  {author} {\bibinfo {author} {\bibfnamefont {L.}~\bibnamefont
  {Agostini}}, \bibinfo {author} {\bibfnamefont {E.}~\bibnamefont {Touber}}, \
  and\ \bibinfo {author} {\bibfnamefont {M.~A.}\ \bibnamefont {Leschziner}},\
  }\href {\doibase 10.1017/jfm.2014.40} {\bibfield  {journal} {\bibinfo
  {journal} {J. Fluid Mech.}\ }\textbf {\bibinfo {volume} {743}},\ \bibinfo
  {pages} {606} (\bibinfo {year} {2014})}\BibitemShut {NoStop}%
\end{thebibliography}%

\end{document}